\documentclass[aps,pra,10pt,twocolumn,groupedaddress,amsmath,amssymb]{revtex4-1}
\usepackage[english]{babel}
\usepackage{graphicx}  
\usepackage{epstopdf}
\usepackage{dcolumn}   
\usepackage{enumerate} 
\usepackage{physics} 
\usepackage{bm}         
\usepackage{verbatim}  
\usepackage{color}
\usepackage[dvipsnames]{xcolor}
\usepackage{diagbox} 

\begin{document}


\title{Quantum metrology in the presence of limited data}
\author{Jes\'{u}s Rubio}
\email{J.Rubio-Jimenez@sussex.ac.uk}
\author{Jacob Dunningham}
\email{J.Dunningham@sussex.ac.uk}
\affiliation{Department of Physics and Astronomy, University of Sussex, Brighton BN1 9QH, UK}
\date{\today}  
   

\begin{abstract}

Quantum metrology protocols are typically designed around the assumption that we have an abundance of measurement data, but recent practical applications are increasingly driving interest in cases with very limited data. In this regime the best approach involves an interesting interplay between the amount of data and the prior information. Here we propose a new way of optimising these schemes based on the practically-motivated assumption that we have a sequence of identical and independent measurements. For a given probe state we take our measurement to be the best one for a single shot and we use this sequentially to study the performance of different practical states in a Mach-Zehnder interferometer when we have moderate prior knowledge of the underlying parameter. We find that we recover the quantum Cram\'{e}r-Rao bound asymptotically, but for low data counts we find a completely different structure. Despite the fact that intra-mode correlations are known to be the key to increasing the asymptotic precision, we find evidence that these could be detrimental in the low data regime and that entanglement between the paths of the interferometer may play a more important role. Finally, we analyse how close realistic measurements can get to the bound and find that measuring quadratures can improve upon counting photons, though both strategies converge asymptotically. These results may prove to be important in the development of quantum enhanced metrology applications where practical considerations mean that we are limited to a small number of trials.

\end{abstract}

\maketitle


\section{Introduction}

Empirical data constitute our primary source of knowledge to construct theories that explain the world around us, and to develop the necessary technologies that help us to accomplish that task. For that reason, how we extract and process that data is a crucial step, and this can be formally captured in quantum systems by the formalism of quantum metrology, a set of techniques that rely on quantum mechanics to extract information about unknown physical quantities from the outcomes of experiments \cite{giovanetti2006review, dunningham2006, paris2009, rafal2015, jesus2017}. 

In practice the quality of this information is restricted by factors such as the number of probes, measurements or repetitions, or by the energy that the experimental arrangement can employ. The latter constraint is particularly relevant for cases where we are interested in studying fragile systems such as atoms, molecules, spin ensembles or biological samples \cite{eckert2007, pototschnig2011, carlton2010, taylor2013, taylor2015, taylor2016, PaulProctor2016}. On the other hand, the number of times that we can interact with the system under study by performing several measurements is always finite and potentially small. This is a possibility that could arise, for instance, in tracking scenarios where we can only have access to a few observations before the object of interest is out of reach, as might be the case for quantum radar \cite{shabir2015, kebei2013, lanzagorta2012} or lidar \cite{lanzagorta2012,wang2016,zhuang2017}.

This situation can be mathematically represented as an optimisation problem where the minimisation of some measure of uncertainty or error for a certain fixed amount of resources informs us about how we should design our experiment so that its performance is optimal. If the uncertainty is based on the square error or can be safely approximated by it, then we can maximise the Fisher information and use the Cram\'{e}r-Rao bound as the figure of merit \cite{paris2009, rafal2015}. This approach is appealing in principle for several reasons. Firstly, bounds for estimating a single parameter derived following this path can always be approached asymptotically provided that we repeat the experiment enough times and that we have certain prior knowledge about the unknown parameter \footnotetext[20]{If the probability model belongs to the exponential family and the estimator is unbiased, then it is possible to saturate the Cram\'{e}r-Rao bound exactly even for a single shot \cite{kolodynski2014, rafal2015, kay1993}. However, in this work we are interested in considering a more general set of scenarios where these restrictions do not necessarily apply, and this generality means that in most cases we can only approach this bound asymptotically.}\cite{rafal2015, jesus2017, braun2018, haase2018jul, Note20}, and this simplifies the optimisation of the error considerably. Furthermore, the Fisher information has a certain fundamental character. In particular, it can be seen as a distinguishability metric \cite{Szczykulska2016} that arises in the expansion of the Bures distance between two infinitesimally close states \cite{paris2009, braun2018}. Moreover, its reciprocal gives us the asymptotic limit for the Bayesian mean square error as a function of the number of repetitions under some fairly general assumptions \footnotetext[22]{It is common to make a distinction between local and global approaches in estimation theory \cite{paris2009, rafal2015}, where the former is associated with the Fisher information and the frequentist interpretation of probabilities and the latter with Bayesian techniques. In that context, it is possible to argue that both approaches are conceptually irreconcilable and that they should be applied to different types of problems \cite{li2018}. However, a conceptually simpler view that also works well is to see Bayesian techniques (in the sense of \cite{jaynes2003}) as the general underlying theory that provides meaningful solutions when well-defined questions are asked, and that, in some cases, these solutions can be approximated using local tools such as the Cram\'{e}r-Rao bound. This perspective was successfully employed in \cite{jesus2017}, and we follow it in this work too.} \cite{jesus2017, Note22}, and this is also the case for other approaches that are more conservative than the Cram\'{e}r-Rao bound too \cite{haase2018may, guta2007}.

Nevertheless, the fact that this technique normally requires many repetitions to be useful is an important drawback to study realistic physical systems such as those previously mentioned. This problem has already been acknowledged in the literature (e.g., in \cite{braun2018, rafal2015, jesus2017,smirne2018}), and several solutions have been proposed. A conceptually simple and straightforward approach consists in using a general measure of uncertainty and estimating how many measurements are needed such that the results predicted by the asymptotic theory are valid, which can always be done numerically \cite{braunstein_gaussian1992, jesus2017}. In addition, we can rely on numerical techniques such as Monte Carlo simulations \cite{braunstein_maxlikelihood1992} or machine learning \cite{lumino2017} to perform the optimisation directly, or can simply examine the behaviour of the system when the number of resources is finite once we have established the asymptotic results \cite{haase2018may}. This was precisely the idea behind the methodology proposed in \cite{jesus2017}, where we analysed the non-asymptotic performance of metrology protocols that had been optimised as if the asymptotic theory were valid, and we explored the structure of the non-asymptotic regime with concrete examples. 

A different possibility is to derive more general lower bounds that are valid in both the asymptotic and the non-asymptotic regimes, such as \cite{tsang2012, tsang2016, liu2016}. Interestingly, this path provides tools that share the computational simplicity of the Cram\'{e}r-Rao bound to some extent, but they also present important limitations. For example, the quantum Ziv-Zakai bound \cite{tsang2012} can recover the asymptotic scaling, but it is not tight in general. The situation improves with the quantum Weiss-Weinstein bound \cite{tsang2016}, since it is asymptotically tight. However, it is not guaranteed that we can saturate this bound in the regime with a finite number of measurements. A similar problem arises with the quantum optimal-bias bound \cite{liu2016}, since by construction it is lower than the Cram\'{e}r-Rao bound and, as we will see, the latter is sometimes lower than the optimal error when it is applied out of its regime of applicability.

This state of affairs motivates the following question: how can we go beyond our current methods and improve our predictions for the optimal performance of our experiments when these operate in the regime of limited data? Here we propose a new method combining analytical and numerical techniques that contributes towards the solution of this problem, and we demonstrate its potential using a Mach-Zehnder interferometer that operates in the regime of limited data and moderate prior knowledge. 

The key idea is to find the measurement scheme predicted by the optimal single-shot mean square error that was originally introduced in \cite{personick1971} and use that measurement in a sequence of repeated experiments. We will show that the bounds that arise from this technique are tight and can be approached in principle both for a single shot (by construction) and in the asymptotic regime of many measurements, since the results predicted by the Fisher information are recovered in the latter case. And while this does not guarantee that our solution will be optimal for a few observations (an adaptive scheme may be better than repeating the same measurement in that case), we will see that having an error that is a function of the number of repetitions where the first point is already tight, and that also tends towards the asymptotically optimal solution as the number of shots grows, is enough to draw conclusions to important questions such as the role of  photon number correlations or the performance of experimentally feasible measurements in the regime of limited data. For instance, we have found an example where the correlations between the paths of the Mach-Zehnder interferometer appear to be particularly useful in this regime, and we have demonstrated that while measuring quadratures and counting photons after the action of a beam splitter are asymptotically equivalent in an ideal scenario, the former measurement scheme is better for a low number of repeated experiments. 

It is interesting to note that a related approach was recently discussed in \cite{esteban2017}, where the authors presented a modification of the quantum Van Trees inequality and used it to construct an adaptive strategy based on an optimal parameter-independent single-shot measurement scheme. Therefore, our work and \cite{esteban2017} are complementary, since we will mainly focus on repeated measurements to connect the optimal single-shot and asymptotic regimes and to explore the regime with a finite number of experiments. Moreover, our results can be seen as a non-trivial generalisation with respect to those that are obtained when the Fisher information is used instead.

The paper is organized as follows. Our method based on single-shot measurements is developed in section \ref{theory}, where we also review the Mach-Zehnder interferometer and the probes that we will use in our calculations. Section \ref{main_results} presents and discusses the bounds that arise from the application of our methodology to optical interferometry, and section \ref{correlations_section} studies the role of intra-mode and inter-mode correlations in the regime of limited data. The effect of changing the prior information is analysed in section \ref{prior_section}, where as expected we recover the predictions of the Fisher information when the prior is very narrow, and we study how to approach our bounds using practical schemes such as photon counting, measurements of quadratures or parity measurements in section \ref{measurements_section}. Finally, a summary of our conclusions and the potential of our proposal for the field of quantum metrology is presented in section \ref{conclusions}.

\section{Framework and strategy}\label{theory}

\subsection{Methodology}

Suppose we have a quantum probe with statistical properties described by the density matrix $\rho_0$. The probe then interacts with a second object characterised by the parameter $\theta$, and this unknown quantity is encoded in the probe state through the unitary transformation $\rho(\theta) = U(\theta)\rho_0 U(\theta)^\dagger$. In order to extract the information about $\theta$ we perform a measurement described by the POVM elements $\lbrace E(n)\rbrace$, where the conditional probability for the outcome $n$ is given by the Born rule $p(n|\theta) = \mathrm{Tr}\left[E(n)\rho(\theta)\right]$. In addition, any extra knowledge that we might have about $\theta$ that is not directly related to the measurement scheme can be encoded in the prior probability $p(\theta)$. 

\begin{figure}
\centering
\includegraphics[trim={0.1cm 0cm 0.4cm 0cm},clip,width=8cm]{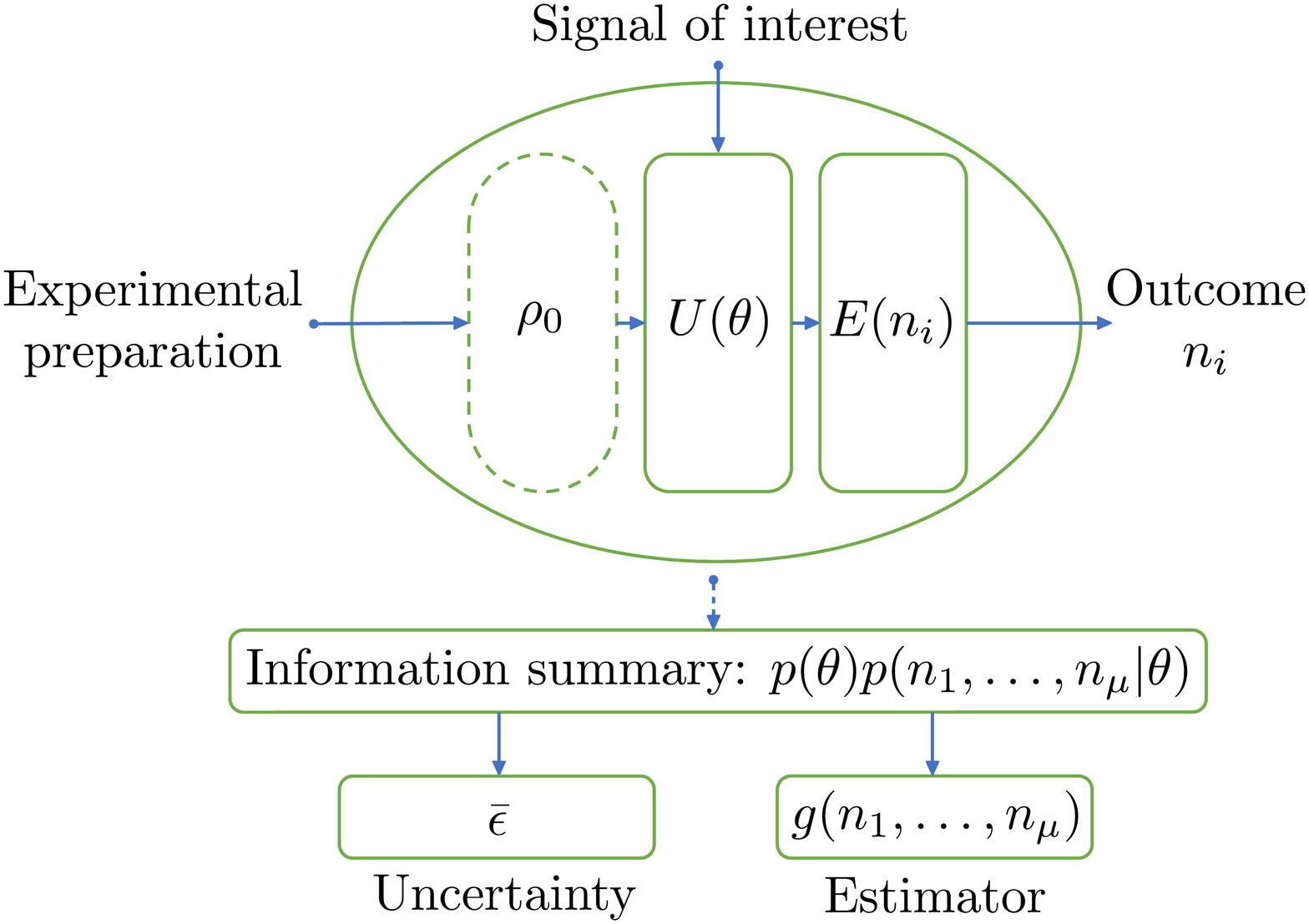}
	\caption{Representation of the extraction of information from a quantum sensor. This process consists of three stages: preparation of the probe state $\rho_0$, parameter encoding $U(\theta)$ and measurement scheme $E(n_i)$. The statistics of the outcome $n_i$ is given by the Born rule, and the protocol is repeated $\mu$ times. Taking also into account any prior information that we may have we can construct an estimator $g(n1,\dots,n_\mu)$ as a function of the experimental outcomes, and assess its performance using some measure of uncertainty $\bar{\epsilon}$.}
\label{prior}
\end{figure}

The joint probability $p(\theta,n)=p(\theta)p(n|\theta)$ contains all the available information about both the experimental outcomes and the unknown parameter. However, to have a more concrete idea about what the value of $\theta$ is we can construct an estimator $g(n)$, which produces an estimate for the parameter as a function of the experimental outcome $n$, and the uncertainty of this procedure can be characterised by the average of some error function $\epsilon\left[g(n),\theta\right]$, that is,
\begin{equation}
\bar{\epsilon} = \int d\theta dn~ p(\theta,n)\epsilon\left[g(n),\theta\right].
\label{singleshoterror}
\end{equation}
As we argued in \citep{jesus2017}, equation (\ref{singleshoterror}) represents the uncertainty on average about the knowledge that we can acquire in principle given the experimental configuration under analysis, and as such it is the suitable quantity to find the optimal strategies for making inferences \footnote{We note that if we wanted to replicate how an experimentalist would update the information provided by a specific experiment, or to simulate the experiment itself, we would need to employ other measures of uncertainty \cite{jesus2017, PaulProctor2016}, as would be the case if we needed to study other quantities such as statistical fluctuations \cite{li2018}}. Note that, at this stage, this is a single-shot quantity. 

Now we focus our attention on the regime of moderate prior knowledge, that is, we are not completely ignorant about the value of the parameter but what we know is not enough to apply the local version of estimation theory. This is motivated by the fact that the prior probability may play a crucial role when the empirical data is limited and, as a consequence, it is the natural regime to study situations where the number of measurements is small. In our case we are going to consider that we know a priori that the parameter is localised somewhere within a domain of width $W_0$, and that this domain is centred around the value $\bar{\theta}$. This state of knowledge can be represented by the uniform density
\begin{equation}
p(\theta) = 1/W_0,~\mathrm{for}~\theta \in [\bar{\theta}-W_0/2, \bar{\theta}+W_0/2],
\label{prior_probability}
\end{equation}
and $p(\theta)=0$ otherwise \footnote{A more realistic prior representing a similar state of knowledge would be, for example, a box-like Gaussian density with a flat peak \cite{braunstein_gaussian1992}. In that sense, the flat prior that we are using is an idealisation.}.

The intermediate regime has been previously explored in the context of optical interferometry \citep{durkin2007, demkowicz2011, jesus2017}. In particular, the method presented in \cite{demkowicz2011} solves the optimisation problem completely using the single-shot error 
\begin{equation}
\bar{\epsilon} = 4\int d\theta dn~ p(\theta,n)\mathrm{sin}^2\left[\frac{g(n) - \theta}{2}\right],
\label{singleshotsinerror}
\end{equation}
which respects the periodic character of the difference of optical phase shifts that we will estimate \cite{holevo2011, helstrom1976, rafal2015}, and it constitutes a particular instance of equation (\ref{singleshoterror}). In principle, we could use the results of \cite{demkowicz2011} and base our analysis in equation (\ref{singleshotsinerror}). However, its extension to the case where many repetitions are considered is still numerically challenging. Instead, in appendix \ref{prior_sinapprox_appendix} we argue that for $W_0 \lesssim 2$ it is meaningful to approximate equation (\ref{singleshotsinerror}) as
\begin{equation}
\bar{\epsilon} \approx \bar{\epsilon}_{\mathrm{mse}} = \int d\theta dn~ p(\theta,n)\left[g(n) - \theta\right]^2,
\label{singleshotmse}
\end{equation}
and we also evaluate the error in the truncation of the Taylor expansion that leads to equation (\ref{singleshotmse}) to show that the main conclusions of this work are not affected by this approximation. In addition, in section \ref{prior_section} we will see that the local regime is not properly recovered until the prior width is $W_0 = 0.1$ or smaller. Hence, this allows us to exploit the simplicity of the mean square error in phase estimation safely within the regime of moderate prior information for $0.1<W_0<2$.

Assuming that the probe state $\rho_0$ and the unitary operator $U(\theta)$ are also known, the next step is to optimise the single-shot mean square error in equation (\ref{singleshotmse}) over all the possible measurement schemes and all the possible estimators. First we note that, according to the proof in \cite{macieszczak2014bayesian}, restricting the possible POVMs to the class of projective measurements does not lead to a loss of optimality in this case. Therefore, we can combine the measurement and the estimator into the observable
\begin{equation}
S = \int dn~g(n) E(n),
\end{equation}
where now $E(n) = \ketbra{n}$, with $\braket{n}{n'}=\delta_{n n'}$, and can rewrite equation (\ref{singleshotmse}) as
\begin{equation}
\bar{\epsilon}_{\mathrm{mse}} = \int d\theta p(\theta) \theta^2 + \mathrm{Tr}\left(\rho S^2 - 2\bar{\rho}S\right),
\label{quantumse}
\end{equation}
with $\rho = \int d\theta p(\theta) \rho(\theta)$ and $\bar{\rho} = \int d\theta p(\theta) \rho(\theta) \theta$. By minimising equation (\ref{quantumse}) with respect to $S$ we finally arrive to \cite{personick1971, macieszczak2014bayesian}
\begin{equation}
\bar{\epsilon}_{\mathrm{mse}} \geqslant \int d\theta p(\theta)\theta^2 - \mathrm{Tr}\left(\bar{\rho}S\right),
\label{singleshot_bound}
\end{equation}
where $S\rho + \rho S  = 2\bar{\rho}$.

The main advantage of this result is that the single-shot optimal strategy can be explicitly constructed from 
\begin{equation}
S = \int ds~s E(s) = \int ds~s \ketbra{s},
\label{singleshot_strategy}
\end{equation}
since this bound is saturated when the projectors $\lbrace \ket{s}\rbrace$ associated with the estimates $\lbrace s\rbrace$ are used as the measurement scheme. In fact, the eigenvalues $\lbrace s\rbrace$ are precisely the estimates given by the mean of the posterior distribution $p(\theta|s)\propto p(\theta) p(s|\theta)$ \cite{personick1971}, which is the classical solution for the optimal estimator \cite{jaynes2003, jesus2017, rafal2015}, and for that reason we will refer to the observable $S$ as the optimal quantum estimator. Moreover, further intuition can be gained by noticing that $\mathrm{Tr}(\rho S) = \int d\theta p(\theta) \theta$ and $\mathrm{Tr}(\bar{\rho} S) = \mathrm{Tr}(\rho S^2)$, so that we can rewrite equation (\ref{singleshot_bound}) as
\begin{equation}
\bar{\epsilon}_{\mathrm{mse}}\geqslant\Delta \theta^2_p - \Delta S^2_{\rho},
\end{equation}
where we have defined the prior uncertainty as
\begin{equation}
\Delta \theta^2_p = \int d\theta p(\theta) \theta^2 - \left[\int d\theta p(\theta) \theta \right]^2
\end{equation}
and 
\begin{equation}
\Delta S^2_\rho = \mathrm{Tr} \left(S^2 \rho \right) - \mathrm{Tr}\left(S \rho \right)^2.
\end{equation}
In words, the uncertainty of our estimation is lower bounded by the difference between the prior variance and the variance of the optimal quantum estimator.

Equation (\ref{singleshot_bound}) was originally discovered and explored in the context of communication theory \cite{personick1969thesis, personick1971, helstrom1976}, and it has been recently used for frequency estimation \cite{macieszczak2014bayesian}. Moreover, a formally similar result emerges in the construction of the quantum Allan variance \cite{chabuda2016allanvariance}. Nevertheless, to the best of our knowledge this result has not been fully exploited to study phase estimation in the regime of limited data and an intermediate prior that we are considering here.

Once the single-shot strategy in equation (\ref{singleshot_strategy}) has been found (we propose a semi-analytical calculation scheme to do this in appendix \ref{numcal}), we proceed to repeat the same optimal experiment $\mu$ times, so that the uncertainty associated with the overall experience is now given by
\begin{equation}
\bar{\epsilon}_{\mathrm{mse}}  = \int d\theta d\boldsymbol{s}~  p(\theta) p(\boldsymbol{s}|\theta)\left[g(\boldsymbol{s}) - \theta\right]^2,
\label{mse_mu_repetitions}
\end{equation}
where $\boldsymbol{s}=(s_1,\dots,s_{\mu})$ is the outcome vector and
\begin{equation}
p(\boldsymbol{s}|\theta) = \prod_{i=1}^\mu\mathrm{Tr}\left[E(s_i)\rho(\theta)\right].
\end{equation}
Moreover, the optimal classical estimator that takes into account the information extracted from all the repetitions is \cite{jaynes2003, jesus2017, rafal2015}
\begin{equation}
g(\boldsymbol{s})=\int d\theta p(\theta|\boldsymbol{s})\theta,
\end{equation}
with $p(\theta|\boldsymbol{s}) \propto p(\theta)p(\boldsymbol{s}|\theta)$. Consequently, the final error is $\bar{\epsilon}_{\mathrm{mse}}  = \int d\boldsymbol{s}~ p(\boldsymbol{s})\epsilon(\boldsymbol{s})$, where  $p(\boldsymbol{s}) = \int d\theta p(\theta)p(\boldsymbol{s}|\theta)$ and $\epsilon(\boldsymbol{s})$ is the variance of the posterior
\begin{equation}
\epsilon(\boldsymbol{s})  = \int d\theta p(\theta|\boldsymbol{s})\theta^2 - \left[\int d\theta p(\theta|\boldsymbol{s})\theta\right]^2.
\end{equation}

The error $\bar{\epsilon}_{\mathrm{mse}}$, which can be numerically calculated as a function of $\mu$ following the three-step scheme discussed in \cite{jesus2017}, is the quantity that we will use to study the low-$\mu$ regime. In other words, our methodology uses numerical simulations and is based on a rigorous foundation provided by an analytical and potentially reachable quantum bound.

Strategies where the same scheme is repeated several times are relevant for any experimental arrangement where we cannot or do not wish to correlate different runs. In that case, it is natural to choose the same optimal single-shot strategy for each individual trial, which also simplifies the complex numerical calculations that are needed to compute $\bar{\epsilon}_{\mathrm{mse}}$ as a function of $\mu$. Admittedly, there are other interesting practical possibilities that emerge when adaptive measurements are allowed \cite{esteban2017, lumino2017}, and while they could be a better choice in some scenarios, adaptive techniques are beyond the scope of this work. Furthermore, from a theoretical perspective we could consider general collective measurements on $\mu$ copies of the same probe. This case is briefly explored for NOON states and a maximum of $\mu=10$ probes in section \ref{measurements_section}, although our main focus is on identical and independent measurements. A discussion about the differences between collective, adaptive and independent measurements is available in \cite{esteban2017}.

\subsection{Physical configuration}\label{employed_states}

The methodology previously introduced is relevant for and can be applied to any unitary estimation problem based on a general mixed probe state where the empirical data is limited and there is a moderate amount of prior knowledge. To illustrate its behaviour, here we will focus on one particular physical configuration.  

Let us consider an interferometer formed by two electromagnetic modes with the same frequency that are modelled by the creation and annihilation operators $a_i^{\dagger}$ and $a_i$, respectively, for $i=1,2$. In addition, for simplicity we assume an ideal situation with pure states.

\begin{table*} [t]
{\renewcommand{\arraystretch}{1.2}
\begin{tabular}{|l|c|c|c|c|c|c|}
\hline
Probe state & $\ket{\psi_0}$  & State parameters  & $\mathcal{Q}$ & $\mathcal{J}$ & $F_q$ & $\mu_{\tau} (\rho)$\\
\hline
\hline
Twin squeezed vacuum state & $S_1(r)S_2(r)\ket{0,0}$ & $r=\mathrm{asinh}\left( 1 \right)$ & $3$ & $0$ & $8$ & $5$ \\ 
\begin{tabular}{@{}l@{}}Twin squeezed cat state (intermediate) \\ Twin squeezed cat state (optimal)\end{tabular} & $\mathcal{N}_{\mathrm{tscs}}\left[S(r)\left(\ket{\alpha}+\ket{-\alpha}\right)\right]^{\otimes 2}$& \begin{tabular}{@{}c@{}}$r=1.103$, $\alpha=1.090$ \\ $r=1.215$, $\alpha=0.9601$\end{tabular} & \begin{tabular}{@{}c@{}}$10.00$ \\ $11.75$\end{tabular}& \begin{tabular}{@{}c@{}}$0$ \\ $0$\end{tabular} & \begin{tabular}{@{}c@{}} $22.00$ \\ $25.49$\end{tabular} & \begin{tabular}{@{}c@{}}$42$ \\ $66$\end{tabular} \\
Squeezed entangled state & $\mathcal{N}_{\mathrm{ses}}\left(\ket{r,0}+\ket{0,r}\right)$ & $r=\log\left(2+\sqrt{3}\right)$ & $9$ & $-0.1$ & $22$ & $45$ \\ 
NOON state & $\left(\ket{N 0}+\ket{0 N}\right)/\sqrt{2}$& $N=2$ &$0$ & $-1$ & $4$ & $116$ \\
Coherent state & $|\alpha/\sqrt{2},-i\alpha/\sqrt{2}\rangle$ & $\alpha = \sqrt{2}$ & $0$ & $0$ & $2$ & $282$ \\ 
\hline
\end{tabular}}
\caption{Properties of the probe states considered in the main text. The state parameters have been chosen such that the mean number of photons is $\bar{n} = 2$. Furthermore, $\mathcal{Q}$ and $\mathcal{J}$ represent the amount of intra-mode and inter-mode correlations in the interferometer, respectively \cite{ShaotaQuesada2015}. Finally, $\mu_{\tau}(\rho)$ indicates the state-dependent number of repetitions that are required for the quantum Cram\'{e}r-Rao bound to be a good approximation to the bounds based on the optimal single-shot strategy in figure \ref{bounds_results}, according to the methodology discussed in \cite{jesus2017} with relative error $\varepsilon_\tau = 0.05$, prior mean $\bar{\theta}=0$ and prior width $W_0 = \pi/2$. Note that $\rho$ includes the information of the initial probe, the encoding of the signal and the prior knowledge.
}
\label{table_summary}
\end{table*}

The benchmark to evaluate the enhancement derived from quantum resources such as entanglement or squeezing will be the coherent state 
\begin{equation}
|\alpha/\sqrt{2},-i\alpha/\sqrt{2}\rangle=\mathrm{exp}\left(-i \frac{\pi}{2} J_x\right)D_1(\alpha)\ket{0,0},
\end{equation}
with $J_x = (a_1^{\dagger}a_2 + a_2 a_1^{\dagger})/2$ and $D_1(\alpha) = \mathrm{exp}(\alpha a_1^{\dagger} - \alpha^{*}a_1)$, while the NOON state $(\ket{N,0} + \ket{0,N})/\sqrt{2}$ will be taken as an example of a definite photon number state that reaches the Heisenberg limit \cite{dowling2008} when enough prior knowledge is available \cite{berry2012, hall2012}. Since many aspects of these two states have been extensively studied in previous works (e.g., in \cite{kolodynski2014, jarzyna2016thesis, hall2012, berry2012}), here we will only highlight those features related to the regime of limited data, and in general we will use them mainly as a reference.

The principal analysis will be dedicated to three experimentally feasible states whose quantum Fisher information is large with respect to the two previous benchmarks \cite{PaulProctor2016}: the twin squeezed vacuum state $\ket{r,r}=S_1(r)S_2(r)\ket{0,0}$, where $S_i(r) = \mathrm{exp}\lbrace[r^{*}a_i^2-r(a_i^{\dagger})^2]/2\rbrace$; the squeezed entangled state $\mathcal{N}_{ses}\left(\ket{r,0}+\ket{0,r}\right)$, where $\mathcal{N}_{ses} = [2+2/\mathrm{cosh}(|r|)]^{-1/2} $; and the twin squeezed cat state $\mathcal{N}_{tscs}\left[S(r)\left(\ket{\alpha}+\ket{-\alpha}\right)\right]^{\otimes 2}$, with $\mathcal{N}_{tscs}=(2+2\mathrm{exp}(-2|\alpha|^2)^{-1/2}$ and $\ket{\alpha} = D(\alpha)\ket{0}$. We recall that the classical Fisher information for a single observation and a given measurement is defined as \cite{rafal2015, kay1993} 
\begin{equation}
F = \int dn~\frac{1}{p(n|\theta)}\left[\frac{\partial p(n|\theta)}{\partial \theta}\right]^2,
\label{classical_fisher}
\end{equation}
and the optimisation of this quantity over all possible POVMs implies that $F \leqslant F_q = \mathrm{Tr}[\rho(\theta)L(\theta)^2]$ \cite{helstrom1976}, where $F_q$ is the quantum Fisher information and the symmetric logarithmic derivative $L(\theta)$ is obtained by solving $L(\theta)\rho(\theta) + \rho(\theta)L(\theta) = 2 \partial \rho(\theta)/\partial \theta$. Moreover, $F$ and $F_q$ do not depend on $\theta$ when the transformation is a unitary that takes the form $U(\theta)=\mathrm{exp}(i H \theta)$, where $H$ is a Hermitian operator. All the protocols that we will study satisfy this condition.

In order to have a fair comparison, the parameters that define the previous states have been chosen such that, on average, all the strategies utilise the same amount of resources (see the third column in table \ref{table_summary}). In particular, $\bar{n} = \bra{\psi_0} (a_1^\dagger a_1+a_2^\dagger a_2) \ket{\psi_0} = 2$ for all $\ket{\psi_0}$. This energy constraint fixes the parameters of all the states except those of the twin squeezed cat state; the parameters of the latter case will be chosen such that the quantum Fisher information is maximum in all the sections of this work except in section \ref{correlations_section}, where we also consider an intermediate scenario. Note that the fact that $\bar{n} = 2$ for all our protocols implies that we are working in the low photon number regime \cite{PaulProctor2016}.

Finally, the unknown parameter $\theta$ represents the difference of phase shifts on the two modes and is encoded using the unitary transformation $U(\theta) = \mathrm{exp}(-i J_z \theta)$, where $J_z = (a_1^{\dagger}a_1 - a_2^{\dagger}a_2)/2$. All the schemes assume that the prior knowledge about this parameter is represented by the probability density in equation (\ref{prior_probability}) with prior width $W_0 = \pi/2 < 2$ and prior mean $\bar{\theta} = 0$.

\section{Quantum bounds in the presence of limited data}\label{main_results}

\begin{figure*}
\centering
\begin{tabular}{l l}
\begin{tabular}{@{}c@{}}\includegraphics[trim={0.2cm 0.1cm 1.2cm 1cm},clip,width=12cm]{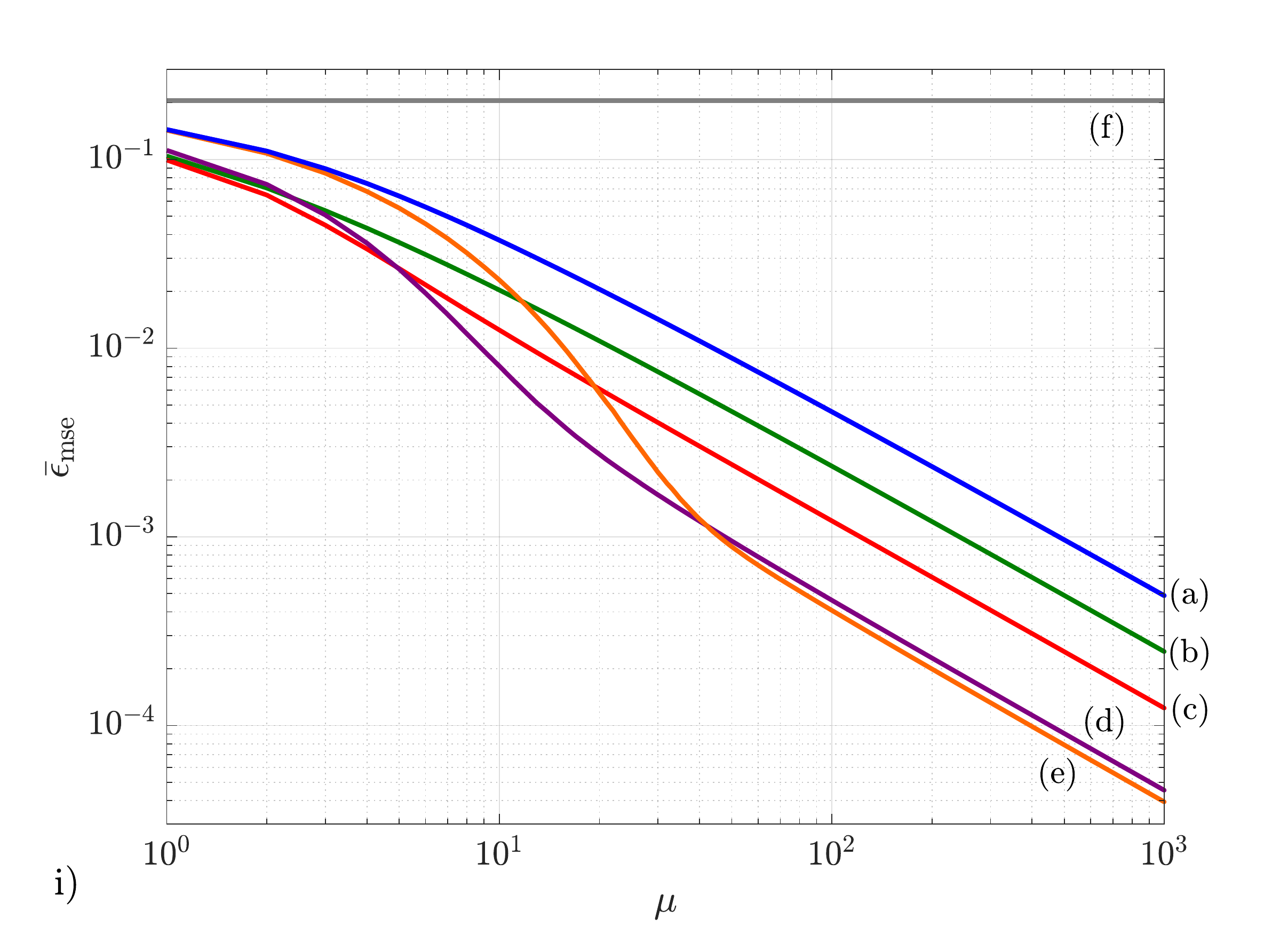}\end{tabular} & \begin{tabular}{@{}c@{}} \includegraphics[trim={0.35cm 0.1cm 1.4cm 1cm},clip,width=6cm]{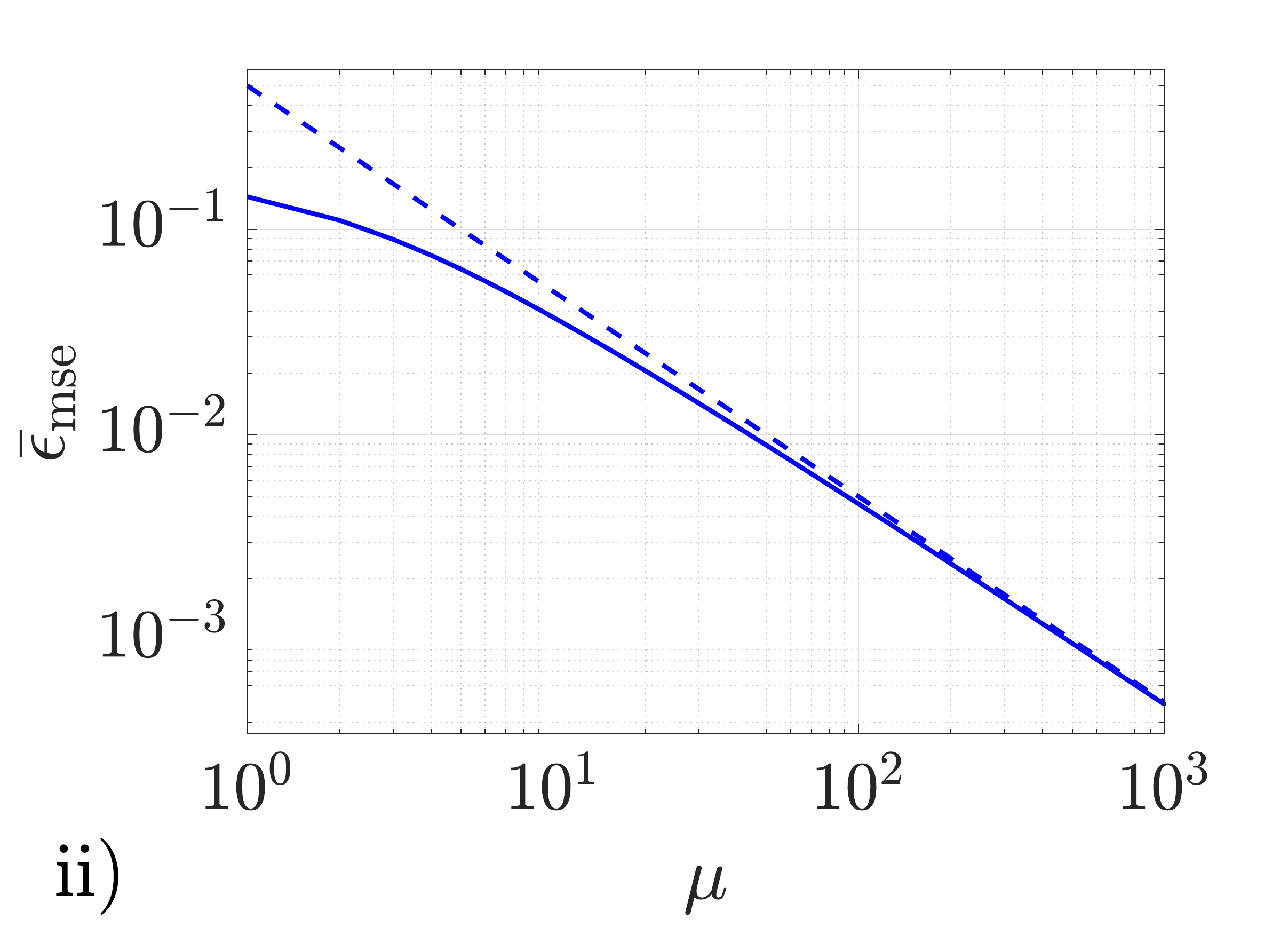} \\ \includegraphics[trim={0.35cm 0.1cm 1.4cm 1cm},clip,width=6cm]{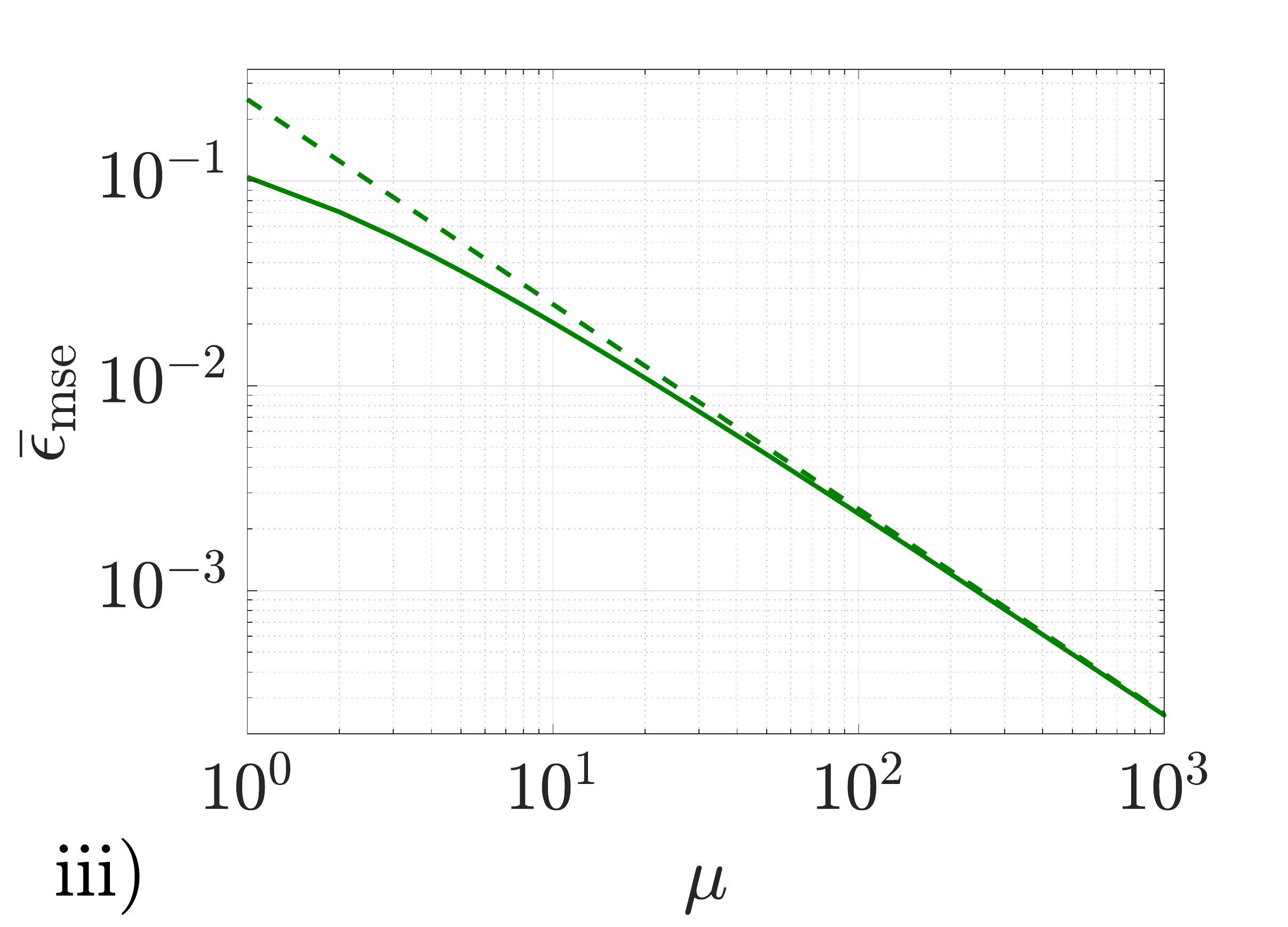}\end{tabular} \\[0pt]
\includegraphics[trim={0.35cm 0.1cm 1.4cm 1cm},clip,width=6cm]{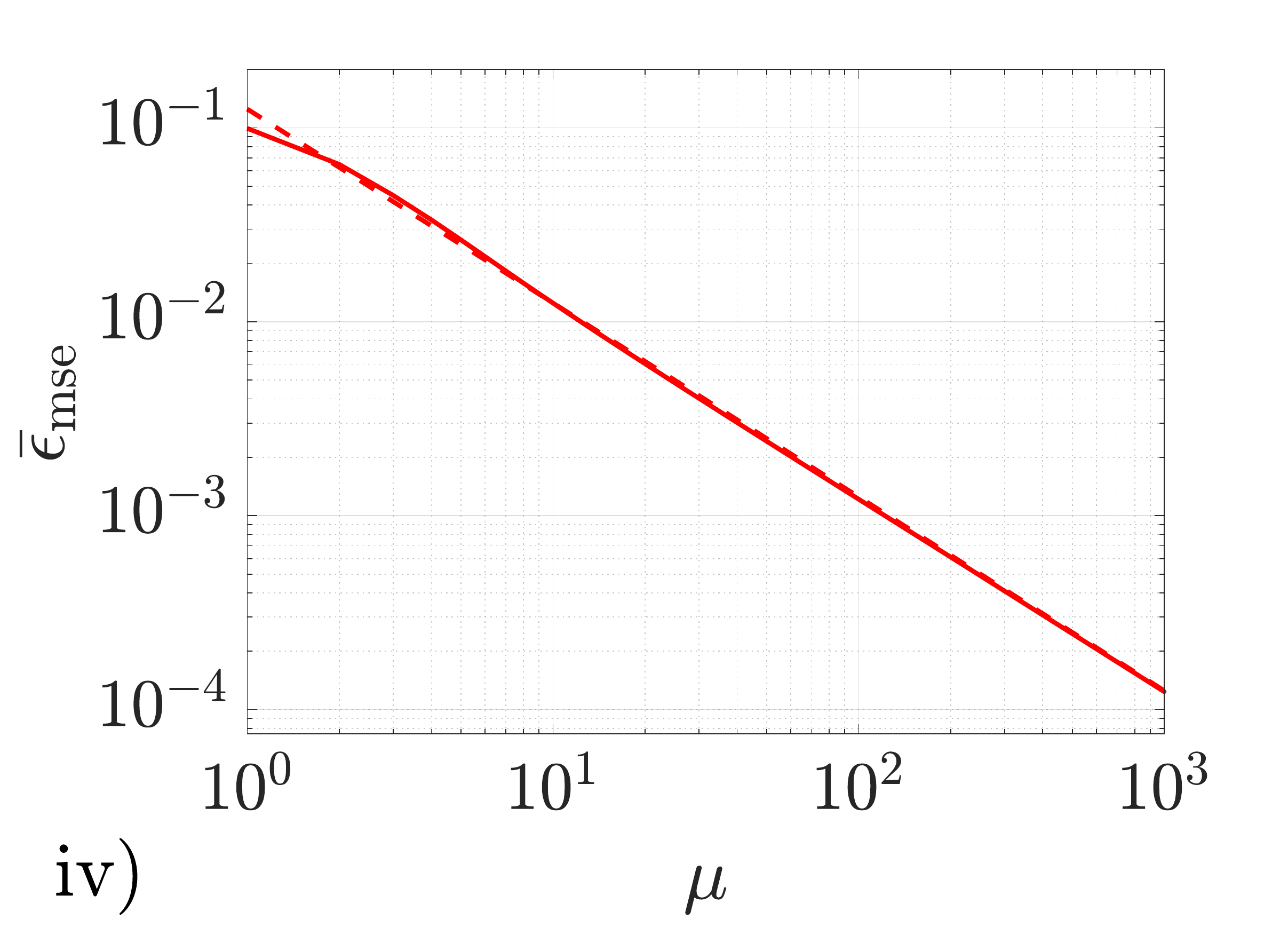}\includegraphics[trim={0.35cm 0.1cm 1.4cm 1cm},clip,width=6cm]{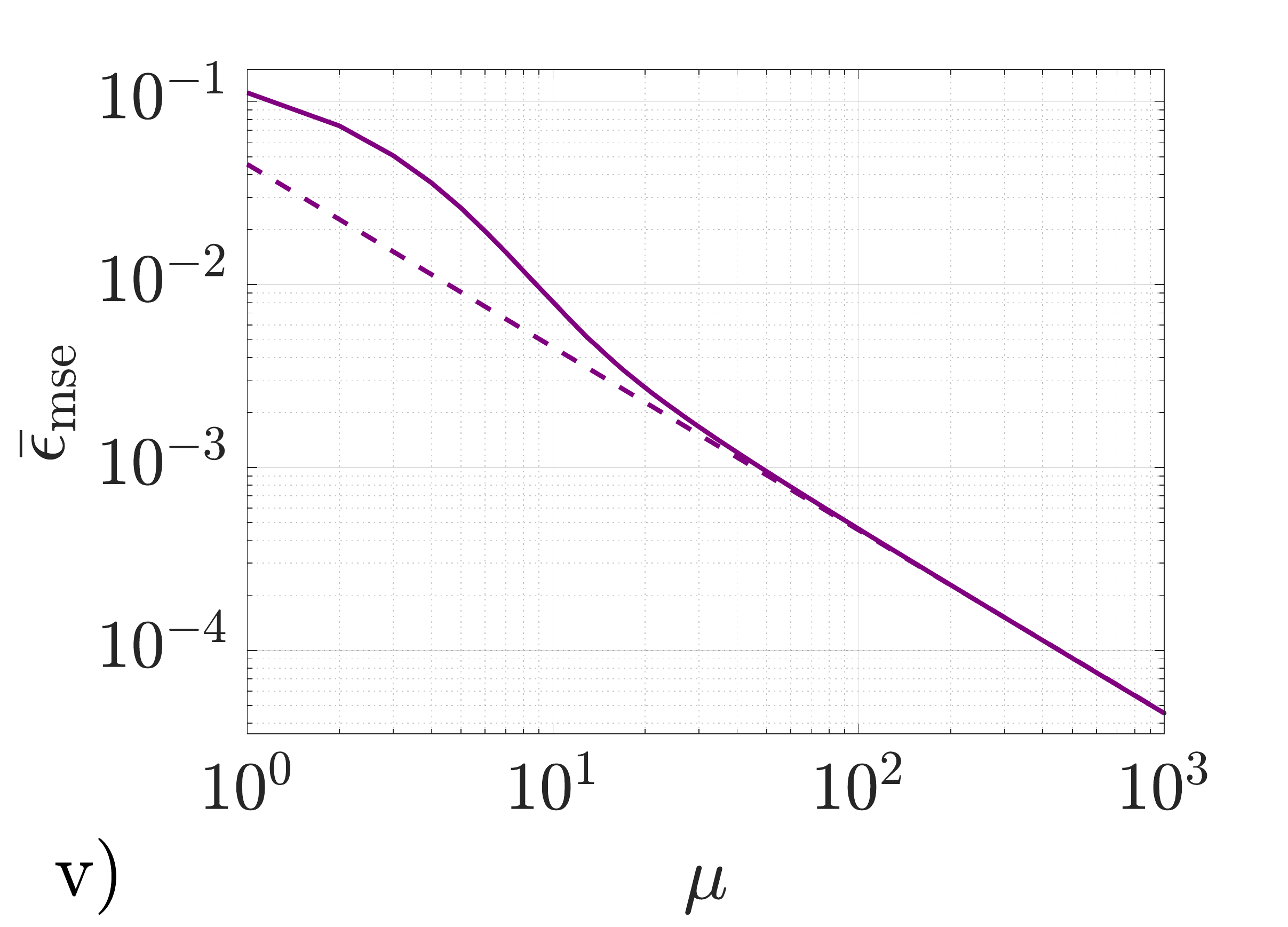} & \includegraphics[trim={0.35cm 0.1cm 1.4cm 1cm},clip,width=6cm]{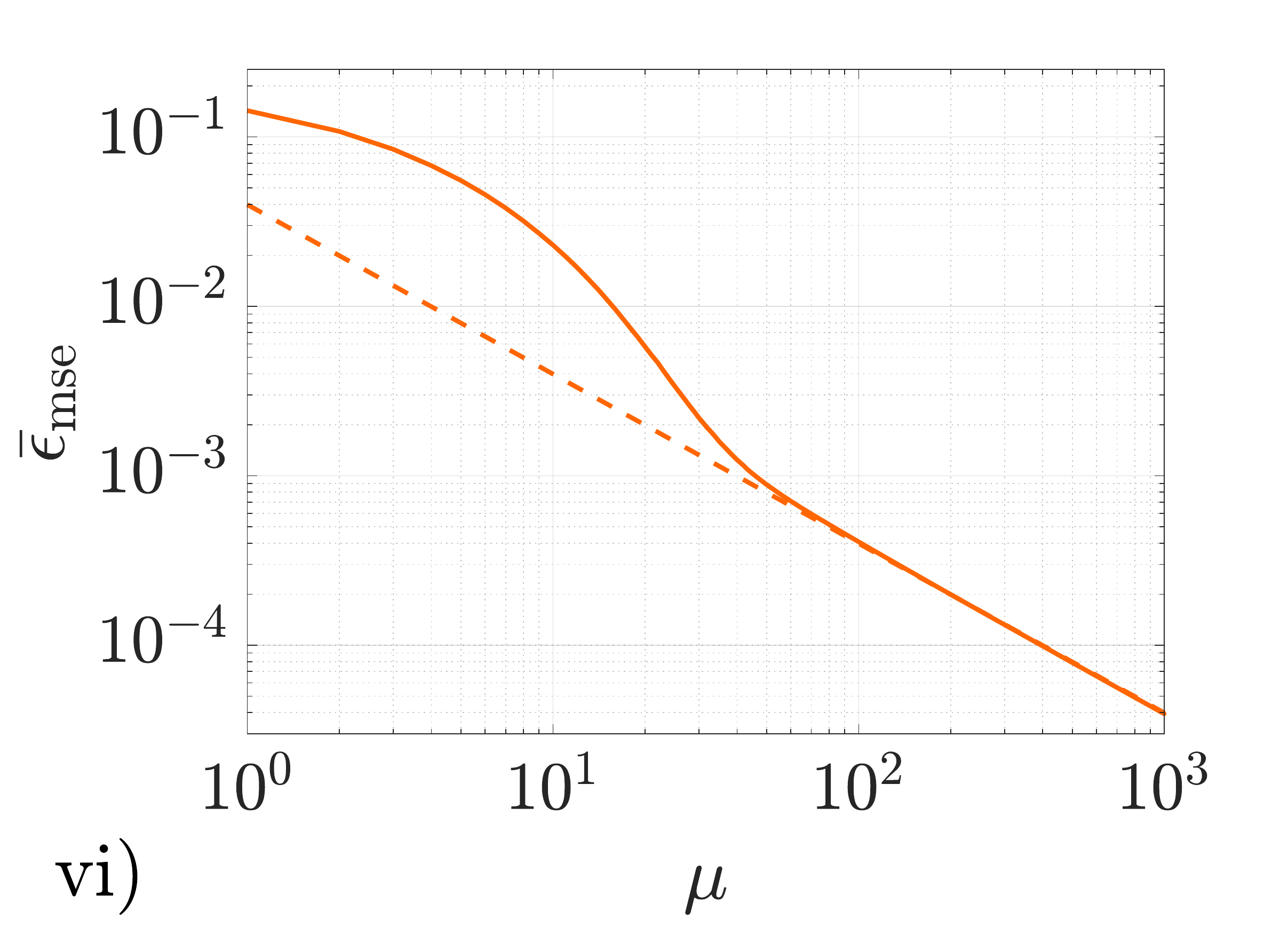}  \\[0pt]
\end{tabular}
\caption{i) Mean square error as a function of the number of repetitions using the optimal single-shot strategy in equation (\ref{singleshot_strategy}) for (a) the coherent state, (b) the NOON state, (c) the twin squeezed vacuum state, (d) the squeezed entangled state, and (e) the twin squeezed cat state, with mean number of photons $\bar{n}=2$, prior mean $\bar{\theta} = 0$ and prior width $W_0 = \pi/2$, while (f) represents the variance of the prior probability; (ii) mean square error based on the optimal single-shot strategy (solid line) and quantum Cram\'{e}r-Rao bound (dashed line) for the same coherent state, (iii) NOON state, (iv) twin squeezed vacuum state, (v) squeezed entangled state and (vi) twin squeezed cat state considered in (i). These graphs constitute the main results of section \ref{main_results}, and their consequences are analysed in the main text.}
\label{bounds_results}
\end{figure*}

The application of the method described in section \ref{theory} to interferometric configurations leads to the results shown in figure \ref{bounds_results}.i, where the mean square error in equation (\ref{mse_mu_repetitions}) is plotted as a function of the number of repetitions for the optical probes previously introduced: (a) coherent state, (b) NOON state, (c) twin squeezed vacuum state, (d) squeezed entangled state and (e) twin squeezed cat state. Let us proceed to analyse the consequences of these graphs.

To start with, figure \ref{bounds_results}.i presents two different regimes. On the one hand, the performance of all the states becomes linear with the number of repetitions in the logarithmic scale when $\mu \gtrsim 10^2$. This is precisely the behaviour that we would expect in the asymptotic regime $\mu \gg 1$, since in that case the mean square error can be approximated by the Cram\'{e}r-Rao bound as $\bar{\epsilon}_{\mathrm{mse}}\approx 1/(\mu F)$ \footnotetext[51]{This approximation relies on the existence of a unique absolute maximum for $p(\boldsymbol{n}|\theta)$ as a function of $\theta$ when $\mu \gg 1$ \cite{jesus2017, cox2000, berry2012}. In other words, the asymptotic likelihood needs to have a single peak within the domain of $\theta$. The largest width employed in this work for the parameter domain is $W_0 = \pi/2$. While this value has been chosen to guarantee that the mean square error is a valid approximation, the analysis of prior information carried out in \cite{jesus2017} demonstrates that, in fact, this choice allows us to approach the Cram\'{e}r-Rao bound using the type of probes considered here. Note that the fact that some states and measurements require a certain amount of prior knowledge to be useful due to the multi-peak structure of their likelihood is ubiquitous in phase-estimation problems \cite{jesus2017, lyons2018, li2018}, although it can be overcome using adaptive techniques \cite{li2018}.} \cite{jesus2017, rafal2015, Note51}, and as such $\mathrm{log}(\bar{\epsilon}_{\mathrm{mse}}) \approx - \mathrm{log}(\mu) - \mathrm{log}(F)$. In this regime we can observe that the graphs of different states do not intersect each other. This property allows us to identify the twin squeezed cat state as the best asymptotic choice, followed by the squeezed entangled state, the twin squeezed vacuum state, the NOON state and, finally, the coherent state, whose performance is the worst. We notice that this is consistent with the findings in \cite{PaulProctor2016}.

On the other hand, the graphs deviate from this logarithmic linear approximation when $1 \leqslant \mu \lesssim 10^2$ and, as a consequence, a non-trivial structure emerges in this part of the plot. This is the non-asymptotic regime of limited data. Since the graphs no longer follow straight lines, they intersect each other, and this implies that the ordering of the states in terms of their performance depends on the number of repetitions. For instance, the twin squeezed vacuum state produces the lowest uncertainty when $1\leqslant\mu<5$, while the squeezed entangled state is the best option when $5<\mu<40$. In addition, the twin squeezed cat state is recovered as the best probe when $\mu>40$, although it practically has the same performance as the coherent state when  $\mu = 1, 2 , 3$. Interestingly, the coherent state is also associated with the largest uncertainty for a low number of trials.

The fact that the strategy leading to the lowest uncertainty can depend on the number of repetitions in a crucial way was already demonstrated in \cite{jesus2017}. However, the results in \cite{jesus2017} were based on a specific measurement scheme (counting photons after the action of a $50$:$50$ beam splitter), while now the bounds are constructed by repeating a single-shot strategy that has been optimised over all possible POVMs. Thus, the results in figure \ref{bounds_results}.i generalise those in \cite{jesus2017} and put the state-dependence behaviour of the non-asymptotic regime on a more solid basis.

For these results to be useful, we need to understand the optimality and saturability of the bounds. The uncertainty for $\mu = 1$ is already optimal by construction and can always be reached in principle for any given state using the single-shot POVM in equation (\ref{singleshot_strategy}). This means that other tools such as the quantum Ziv-Zakai bound \cite{tsang2012} and the quantum Weiss-Weinstein bound \cite{tsang2016} will necessarily produce less tight single-shot results whenever their value is different from the solution found here. 

\begin{figure}[ht]
\centering
\includegraphics[trim={2.25cm 0.25cm 1.3cm 1cm},clip,width=9.05cm]{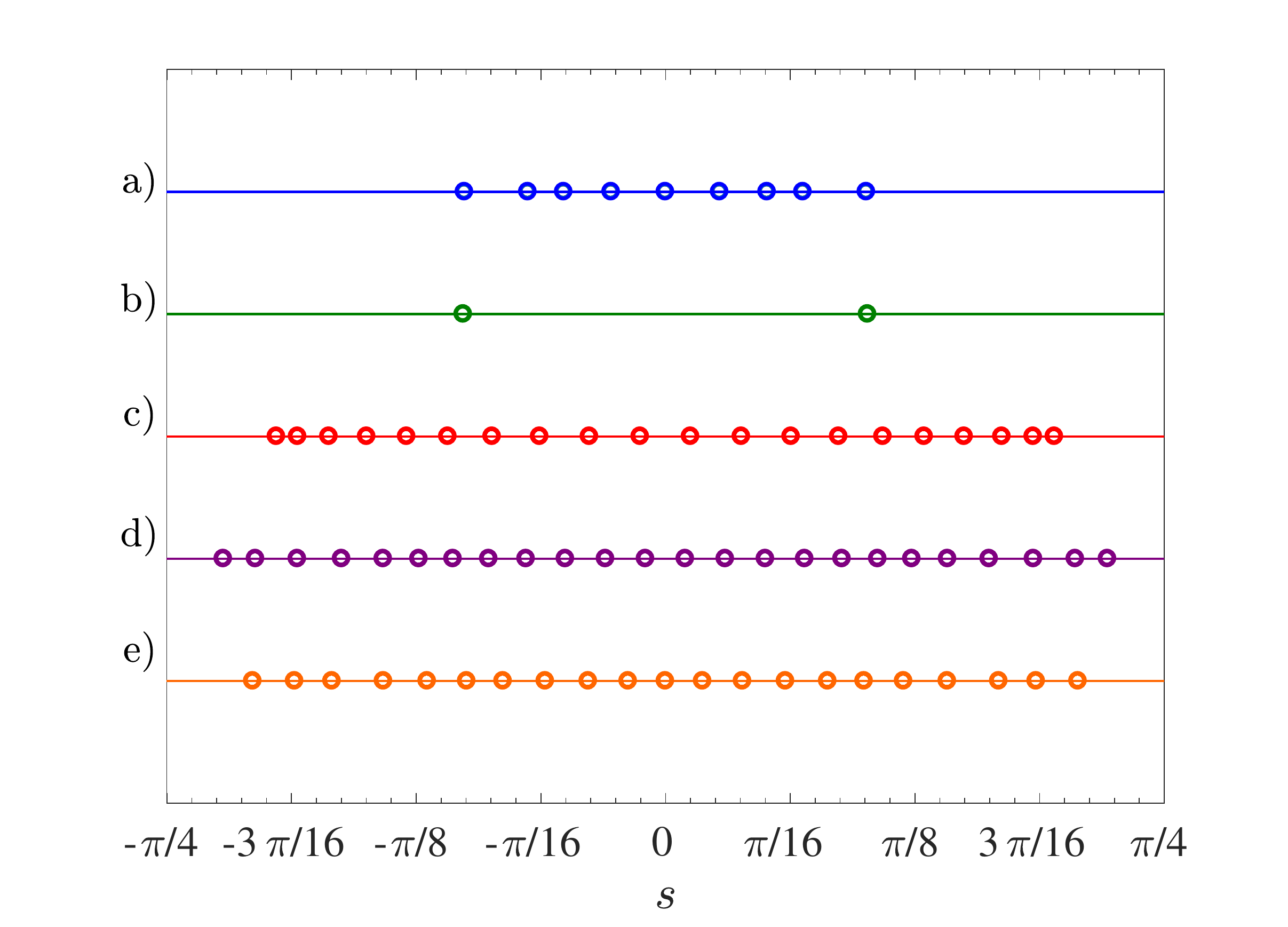}
	\caption{Spectrum of the optimal quantum estimator $S$ for (a) the coherent state, (b) the NOON state, (c) the twin squeezed vacuum state, (d) the squeezed entangled state, and (e) the twin squeezed cat state, with $\bar{n}=2$, $\bar{\theta} = 0$ and $W_0 = \pi/2$. The details of this calculation can be found in appendix \ref{numcal}.}
\label{bayes_spectra}
\end{figure}

Furthermore, figures \ref{bounds_results}.ii - \ref{bounds_results}.vi show how our results for each state approach the quantum Cram\'{e}r-Rao bound asymptotically, that is, $\bar{\epsilon}_{\mathrm{mse}} \approx 1/(\mu F_q)$  when $\mu \gg 1$. Taking into account that  the bounds for a large number of trials that can be constructed using the quantum Cram\'{e}r-Rao bound are fundamental, we conclude that our bounds are also optimal in this limit. As a result, if we work in the regime of intermediate prior knowledge and $\rho(\theta)$ and $p(\theta)$ are given, then the scheme developed in section \ref{theory} is optimal both for a single shot and a large number of trials. Moreover, it is also optimal for any number of trials if we exclude the possibility of having adaptive measurements and focus on identical and independent experiments.

To quantify the number of repetitions that are needed to reach this asymptotic regime where our methods are not longer required we can follow \cite{jesus2017}, construct the relative error 
\begin{equation}
\varepsilon_\tau = \frac{|\bar{\epsilon}_{\mathrm{mse}}(\mu_{\tau}) - 1/(\mu_{\tau} F_q)|}{\epsilon_{\mathrm{mse}}(\mu_{\tau})}
\label{threshold}
\end{equation}
and select $\mu_{\tau}$ after imposing that $\varepsilon_\tau \approx 0.05$ for each state. According to the results of this calculation, which are summarised in the last column of table \ref{table_summary}, the uncertainty for the twin squeezed vacuum state agrees with the prediction of the quantum Cram\'{e}r-Rao bound when the number of trials is as low as $\mu_{\tau} = 5$. Therefore, in this case the asymptotic theory mostly gives the right answer. However, the squeezed entangled state and the twin squeezed cat state require $\mu_{\tau}=45$ and $\mu_{\tau}=66$, respectively, and the quantum Cram\'{e}r-Rao bound overestimates the performance of these probes in the regime of limited data because the graphs of our bounds are higher (figures \ref{bounds_results}.v and \ref{bounds_results}.vi). We note that it is in scenarios of this type where we could not extract useful information from the quantum optimal-bias bound derived in \cite{liu2016}, since for a flat prior this quantity is always lower than the quantum Cram\'{e}r-Rao bound by construction. Finally, the NOON state needs $\mu_{\tau}=116$ and the coherent state requires $\mu_{\tau}=282$, but the Cram\'{e}r-Rao bound prediction underestimates the precision of these protocols when $\mu$ is low. It is interesting to observe that the chosen probes exemplify the three basic behaviours that we could expect to find in the non-asymptotic regime, that is, that the Cram\'{e}r-Rao bound is lower, higher or approximately equal to the Bayesian mean square error.

Although the numerical character of the previous results does not reveal the structure of the optimal measurement scheme associated to each state, it is possible to gain some intuition by studying the optimal quantum estimator $S$ in equation (\ref{singleshot_strategy}). For the NOON state we have calculated this operator analytically in appendix \ref{noon_analytical}, finding that its projectors are simply
\begin{eqnarray}
\ket{s_1} &=& \frac{1}{\sqrt{2}} (i\ket{2,0}+\ket{0,2}), 
\nonumber \\
\ket{s_2} &=& \frac{1}{\sqrt{2}} (\ket{2,0}+i\ket{0,2}),
\label{noon_projectors}
\end{eqnarray}
and that its eigenvalues are $s_1=-1/\pi$ and $s_2=1/\pi$, which represent the Bayesian estimates for $\theta$. In section \ref{measurements_section} we will construct physical measurements that realise these projectors exactly. In addition, it is important to note that, while this spectrum of estimates is discrete and the difference of phase shifts $\theta$ is a continuous variable, in \cite{alfredo1996} it was shown that this behaviour is not contradictory due to the existence of an ultimate quantum limit to the uncertainty in phase estimation.

On the other hand, a fully analytical calculation of $S$ for the indefinite photon number states is more challenging, and the numerical projectors that arise from the calculation scheme proposed in appendix \ref{numcal}, which have been used to find the results in figure \ref{bounds_results}, are difficult to visualise. Nevertheless, we can still provide a partial characterisation of the single-shot strategies through the spectrum of the optimal quantum estimator for different states. A numerical approximation of these spectra has been represented in figure \ref{bayes_spectra} for the coherent state, the twin squeezed vacuum state, the squeezed entangled state and the twin squeezed cat states, which shows their Bayesian estimates distributed within the parameter domain $[\bar{\theta}-W_0/2,\bar{\theta}+W_0/2]$\footnotetext[53]{The solution to the Sylvester equation $S\rho + \rho S=2\bar{\rho}$ is only unique when the spectra of $\rho$ and $-\rho$ are disjoint \cite{bhatia1997} and, in fact, the quantity $\mathrm{Tr}(\bar{\rho}S) = \mathrm{Tr}(\rho S^2)$ appearing in the single-shot bound in equation (\ref{singleshot_bound}) only depends on the terms associated with the support of $\rho$. We note that the particular number of estimates represented in the approximated spectra of figure \ref{bayes_spectra} for each indefinite photon number state depends on the numerical truncation of this support, which in our case assumes that an eigenvalue of $\rho$ is non-zero when its value is higher than $\sim 10^{-12}$. See Appendix \ref{numcal} for more details about the numerical approximations employed in this work.} \cite{Note53}.

We finish this analysis by noting that both the projectors $\lbrace \ket{s} \rbrace$ and the estimates $\lbrace s \rbrace$ depend on the specific shape of the prior probability $p(\theta)$. Interestingly, in our case we have verified numerically that while the results change with $W_0$, they do not depend on $\bar{\theta}$. Nonetheless, in section \ref{measurements_section} we will see that this is no longer true for measurement schemes different from the optimal single-shot strategy.

\section{The role of intra-mode and inter-mode correlations for a low number of repetitions}\label{correlations_section}

In the context of optical interferometry there are two types of correlations that are relevant for quantum metrology: the intra-mode correlations quantified by the Mandel $\mathcal{Q}$-parameter, which for the path-symmetric states that we are considering here can be written as \cite{ShaotaQuesada2015}
\begin{equation}
\mathcal{Q} = \frac{4\langle (a_1^{\dagger}a_1)^2 \rangle - \bar{n}^2 -2\bar{n}}{2\bar{n}} = \frac{4\langle (a_2^{\dagger}a_2)^2 \rangle - \bar{n}^2 -2\bar{n}}{2\bar{n}},
\label{mandel}
\end{equation}
where we are using the notation $\langle \Box \rangle = \bra{\psi_0} \Box \ket{\psi_0}$, and the inter-mode correlations quantified by \cite{ShaotaQuesada2015}
\begin{equation}
\mathcal{J} = \frac{\langle a_1^{\dagger}a_1 a_2^{\dagger}a_2\rangle-\bar{n}^2/4}{\Delta(a_1^{\dagger}a_1)\Delta(a_2^{\dagger}a_2)},
\label{entanglement}
\end{equation}
where we have also incorporated the fact that the states are path-symmetric. 

These quantities play a crucial role in the regime where $\bar{\epsilon}_{\mathrm{mse}} \approx 1 /(\mu F_q)$ because the quantum Fisher information for path-symmetric pure states can be rewritten as $F_q = \bar{n}(1+\mathcal{Q})(1-\mathcal{J})$ \cite{ShaotaQuesada2015, knott2016local}. Therefore, we can control the asymptotic performance by changing $\mathcal{Q}$ and $\mathcal{J}$. Recalling that $-1\leqslant\mathcal{Q}<\infty$ and $-1\leqslant\mathcal{J}\leqslant 1$, optimising the performance amounts to increasing the intra-mode correlations as much as possible, since path entanglement can only improve the precision by a factor of $2$ at most. To verify that the asymptotic part of figure \ref{bounds_results}.i is consistent with this way of proceeding we have calculated the amount of intra-mode and inter-mode correlations and the quantum Fisher information for each state \footnote{An analytical calculation of these quantities for coherent, NOON and twin squeezed vacuum states is available in \cite{ShaotaQuesada2015}, while the results for squeezed entangled and twin squeezed cat states can be found in \cite{PaulProctor2016}.}, and the results can be found in the fourth, fifth and sixth columns of table \ref{table_summary}, respectively. As expected, the twin squeezed cat state, which was found to be the asymptotically optimal choice, has the largest values for $F_q$ and $\mathcal{Q}$ among the states that we are studying.

On the other hand, we have also demonstrated that this state is not better than a coherent state when $\mu \sim 1$, in spite of the fact that for the coherent state we have $\mathcal{Q} = 0$ and $\mathcal{J}=0$, and that the other three probes perform better in the low trial number regime. This already supports the idea that the clear role that photon number correlations play asymptotically is not preserved when $\mu$ is low, something that was suggested in \cite{jesus2017} using a specific POVM. While it is not currently possible to find a rigorous relationship between uncertainty and correlations that is also valid in the regime of limited data because an analytical expression for $\bar{\epsilon}_{\mathrm{mse}}(\mu)$ is not available, we can still exploit the methodology introduced in section \ref{theory} to further explore this idea.

\begin{figure}[t]
\centering
\includegraphics[trim={0.2cm 0.1cm 1.3cm 1cm},clip,width=9cm]{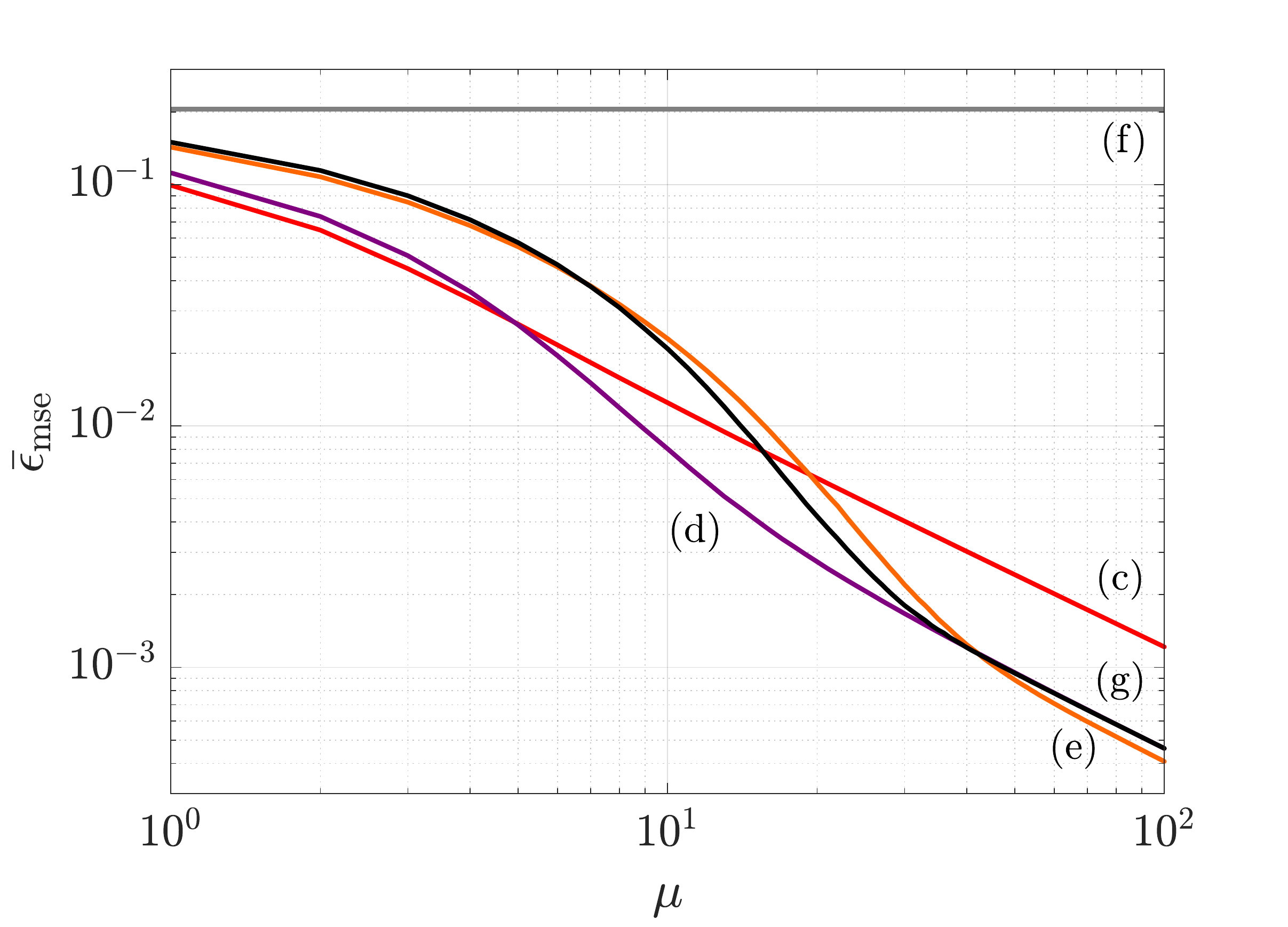}
	\caption{Mean square error as a function of the number of repetitions using the optimal single-shot strategy in equation (\ref{singleshot_strategy}) for (c) the twin squeezed vacuum state with $\mathcal{Q}=3$ and $\mathcal{J} = 0$, (d) the squeezed entangled state with $\mathcal{Q}=9$ and $\mathcal{J} = -0.1$, (e) the twin squeezed cat state with $\mathcal{Q}=11.75$ and $\mathcal{J} = 0$, and (e) the twin squeezed cat state with $\mathcal{Q}=10.00$ and $\mathcal{J} = 0$, where $\mathcal{Q}$ and $\mathcal{J}$ quantify the intra-mode and inter-mode correlations, and having $\bar{n}=2$, $\bar{\theta} = 0$ and $W_0 = \pi/2$, while (f) represents the variance of the prior probability.}
\label{correlations_figure}
\end{figure}

First we note that the twin squeezed cat state can be seen as a family of states defined in terms of the parameters $r$ and $\alpha$. Since this state is separable with respect to the arms of the interferometer, $\mathcal{J} = 0$, and as such we are free to choose different combinations of $r$ and $\alpha$ to control the Mandel $\mathcal{Q}$-parameter while keeping $\bar{n}=2$ and $W_0=\pi/2$ unchanged. The particular instance of the twin squeezed cat family with $\mathcal{Q} = 11.75$ and $F_q = 25.49$ considered until now is the optimal choice after maximising $F_q$ numerically. A second example with $\mathcal{Q} = 10.00$ and $F_q = 22.00$ has been included in table \ref{table_summary} to represent the intermediate case. In addition, the twin squeezed vacuum state is recovered within the twin squeezed cat family when we choose $\alpha = 0$ \cite{PaulProctor2016}, and for this state we have that $\mathcal{Q} = 3$ and $F_q = 8$. 

\begin{table*} [t]
{\renewcommand{\arraystretch}{1.2}
\begin{tabular}{|l|c|c|c|c|c|}
\hline
\diagbox{Probe state ~~~~~~~~~~}{$\mu\cdot\bar{\epsilon}_{\mathrm{mse}}(\mu, W_0)$} & $\mu=1$, $W_0=\pi/2$ & $\mu=1$, $W_0=\pi/3$ & $\mu=1$, $W_0=\pi/4$ & $\mu=1$, $W_0=0.1$ & $\mu \gg 1$ \\
\hline
\hline 
Twin squeezed vacuum state & $9.93\cdot 10^{-2}$ & $5.83\cdot 10^{-2}$ & $3.81\cdot 10^{-2}$ & $8.28\cdot 10^{-4}$ & $1.25\cdot 10^{-1}$  \\ 
Twin squeezed cat state (intermediate) & $1.50\cdot 10^{-1}$ & $6.48\cdot 10^{-2}$ & $3.61\cdot 10^{-2}$ & $8.19\cdot 10^{-4}$ & $4.55\cdot 10^{-2}$  \\ 
Twin squeezed cat state (optimal) & $1.42\cdot 10^{-1}$ & $7.10\cdot 10^{-2}$ & $4.11\cdot 10^{-2}$ & $8.17\cdot 10^{-4}$ & $3.92\cdot 10^{-2}$  \\ 
Squeezed entangled state & $1.12\cdot 10^{-1}$ & $5.61\cdot 10^{-2}$ & $3.47\cdot 10^{-2}$ & $8.19\cdot 10^{-4}$ & $4.55\cdot 10^{-2}$   \\ 
NOON state & $1.04\cdot 10^{-1}$ & $6.47\cdot 10^{-2}$ & $4.21\cdot 10^{-2}$ & $8.31\cdot 10^{-4}$ & $2.5\cdot 10^{-1}$  \\
Coherent state & $1.44\cdot 10^{-1}$ & $7.71\cdot 10^{-2}$ & $4.66\cdot 10^{-2}$ & $8.33\cdot 10^{-4}$ & $5\cdot 10^{-1}$   \\ 
\hline
\end{tabular}}
\caption{Optimal single-shot mean square error with different prior widths for the states considered in the main text and asymptotic performance for $\mu\gg1$. The state parameters are those indicated in table \ref{table_summary}. We notice that the asymptotic ordering of probe states and the ordering for $W_0=0.1$ and a single shot are identical, which implies that the local regime is achieved for such a prior width.}
\label{prior_effect_summary}
\end{table*}
 
Next we examine the mean square errors associated with the optimal case, the intermediate case and the twin squeezed vacuum from the previous family. Their graphs are represented in figure \ref{correlations_figure} and labelled respectively as (e), (g) and (c). If we compare the optimal and intermediate states first, we see that a larger amount of intra-mode correlations is associated with a larger number of repetitions needed to reach the asymptotic regime, since the former state requires $\mu_\tau=66$ and the latter $\mu_\tau=42$ (see table \ref{table_summary}). Furthermore, by comparing the form of the graphs (e) and (g) in figure \ref{correlations_figure} for these two states we can observe that the transition from the non-asymptotic regime to the asymptotic regime is associated  with a larger uncertainty for the optimal twin squeezed cat state for which $\mathcal{Q}$ is also larger. Finally, the graph (c) shows that the twin squeezed vacuum state, which has the smallest $\mathcal{Q}$, performs worse than the two previous cases asymptotically, while its error is the lowest when $1\leqslant\mu \lesssim 10$. In other words, for this family of states there seems to be a trade-off between the performances in the asymptotic and non-asymptotic regimes that is associated with changes in $\mathcal{Q}$, which in practice would imply that increasing the amount of intra-mode correlations blindly can lead to high-uncertainty schemes in the regime of limited data. Moreover, we note that this conclusion is consistent with the related analysis in \cite{tsang2012} for the Rivas-Luis state \cite{rivas2012} based on the quantum Ziv-Zakai bound, which demonstrated that if a certain parameter is modified such that the Fisher information increases arbitrarily, then the error cannot deviate substantially from the prior variance unless the number of trials is very large.

Since increasing $\mathcal{Q}$ seems to be detrimental to the performance of our probes when the number of repetitions is  low, the next natural step is to investigate whether path entanglement could be useful in this regime. Including in our analysis the squeezed entangled state with $\mathcal{Q} = 9$ and $\mathcal{J}=-0.1$, which is labelled as (d) in figure \ref{correlations_figure}, we can see that this state converges asymptotically to the performance associated with the intermediate case of the twin squeezed cat family (g), that is, both probes have the same Fisher information. However, the graph of the squeezed entangled state presents a smaller curvature and a lower uncertainty when $\mu < 30$. The key aspect that distinguishes these two probes is that the squeezed entangled state has a lower amount of intra-mode correlations and a certain amount of beneficial path entanglement, which suggests that inter-mode correlations have helped to improve the precision in the non-asymptotic regime while keeping a large Fisher information. Hence, we conclude that path entanglement could be considerably more relevant in schemes that need to be optimised for a low number of trials than it is in the asymptotic regime. 

Despite these surprising results, we must acknowledge that our analysis is centred on a particular set of states, and that other schemes based on different states could show different properties \footnote{Furthermore, it is reasonable to expect that other schemes that allow other types of correlations behave differently too. For instance, in \cite{smirne2018} the authors showed that allowing entanglement between a finite number of probes in a frequency estimation protocol can lead to a less precise strategy.}. Therefore, the existence of a more general relationship between the number of trials and the usefulness of photon number correlations in interferometry for a given prior is an open question. 

\section{The effect of the prior information}\label{prior_section}

In a wide set of inference problems that includes the metrology scenarios presented here, the importance of the prior information depends on the number of shots. In particular, we know that the prior becomes less important as we increase the number of repetitions \cite{jesus2017, cox2000}. This implies that, as we argued in section \ref{theory}, the prior probability is going to play an important role for making inferences if only a few experimental shots are possible. In that scenario it is crucial then to establish how different states of prior knowledge may affect the overall performance of a given metrology scheme.

\begin{figure*}[t]
\centering
\includegraphics[trim={0.1cm 0.1cm 1.2cm 0.2cm},clip,width=9.1cm]{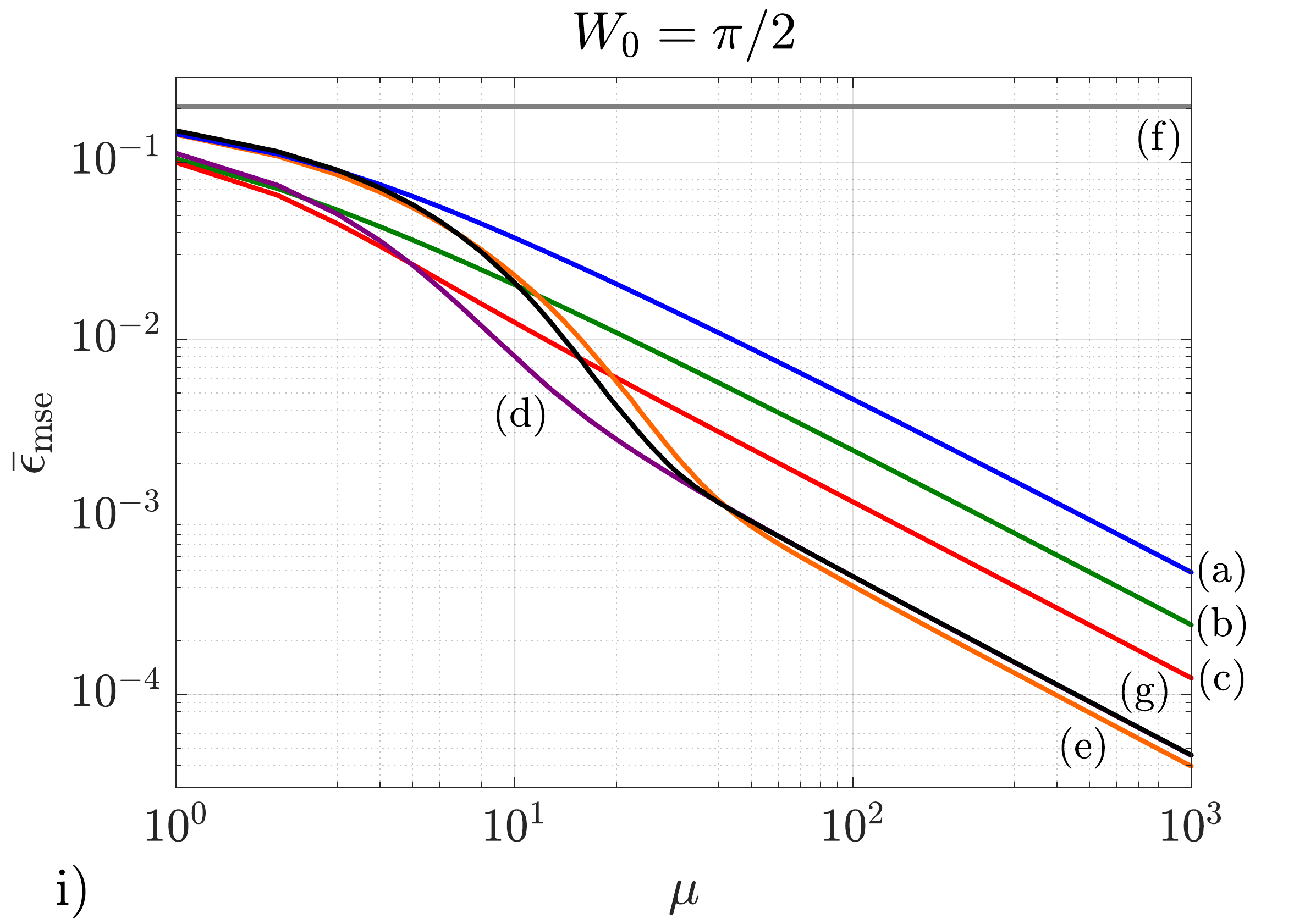}\includegraphics[trim={0.1cm 0.1cm 1.3cm 0.2cm},clip,width=9.1cm]{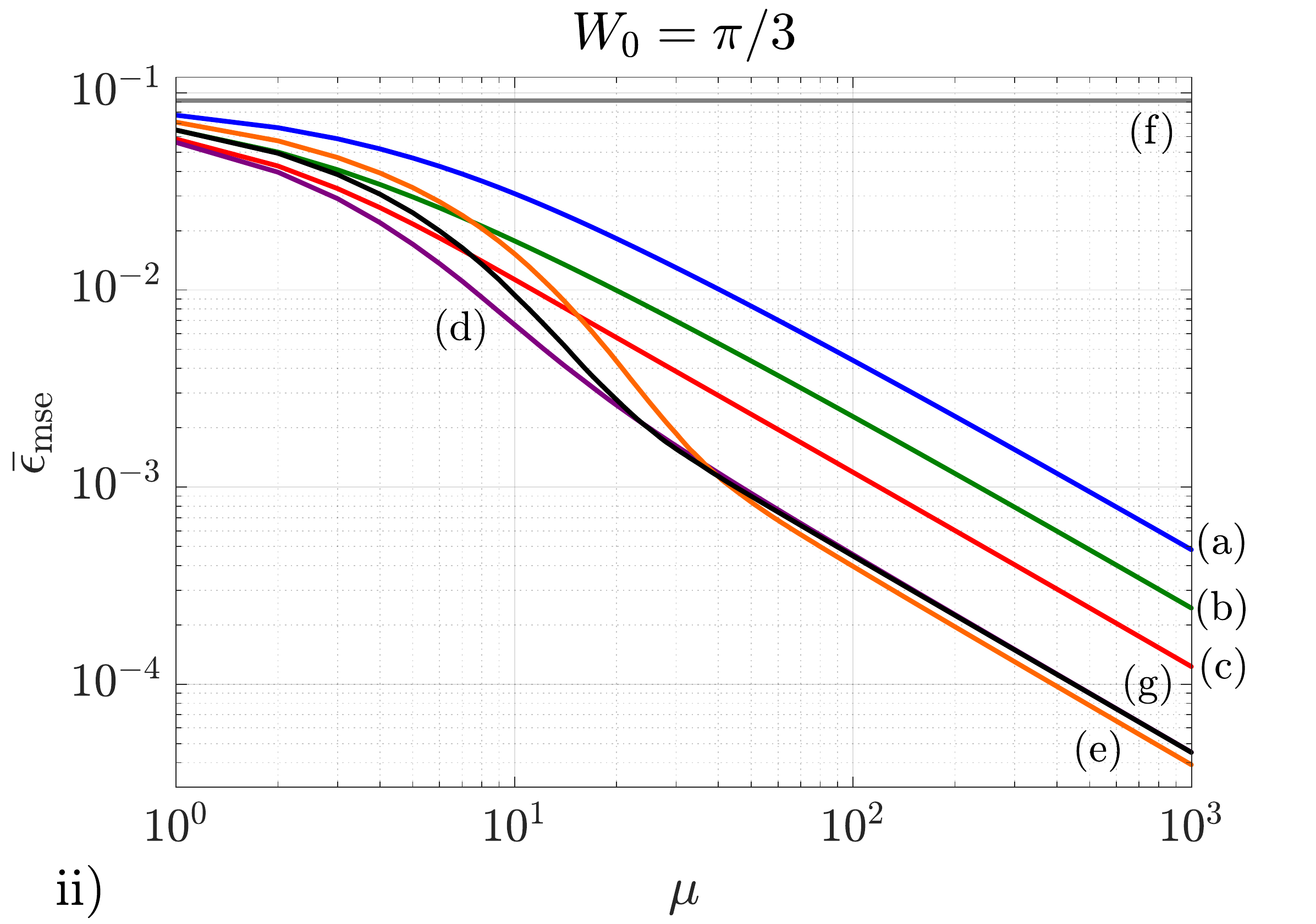}
\includegraphics[trim={0.1cm 0.1cm 1.2cm 0cm},clip,width=9.1cm]{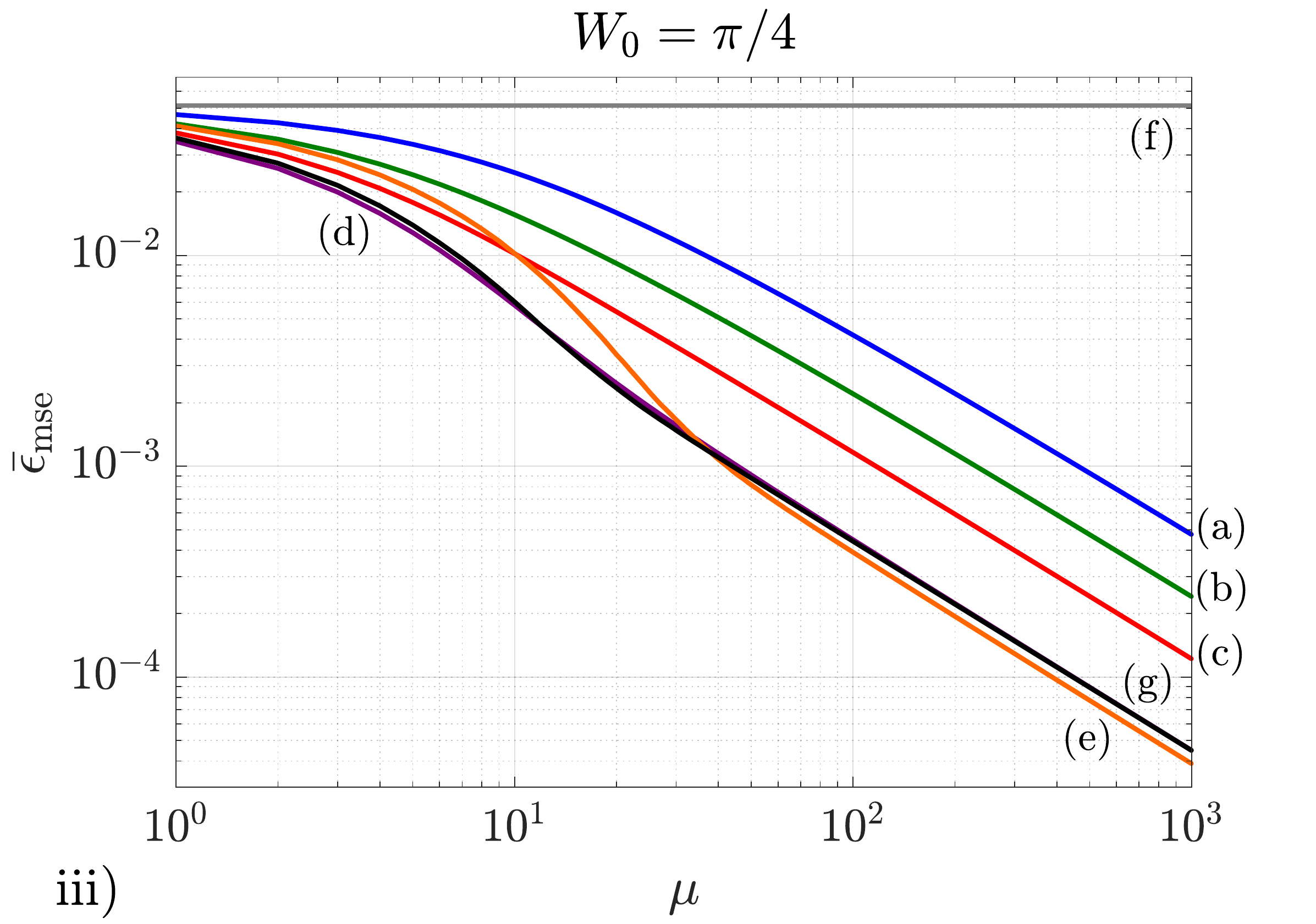}\includegraphics[trim={0.1cm 0.1cm 1.3cm -0.5cm},clip,width=9.1cm]{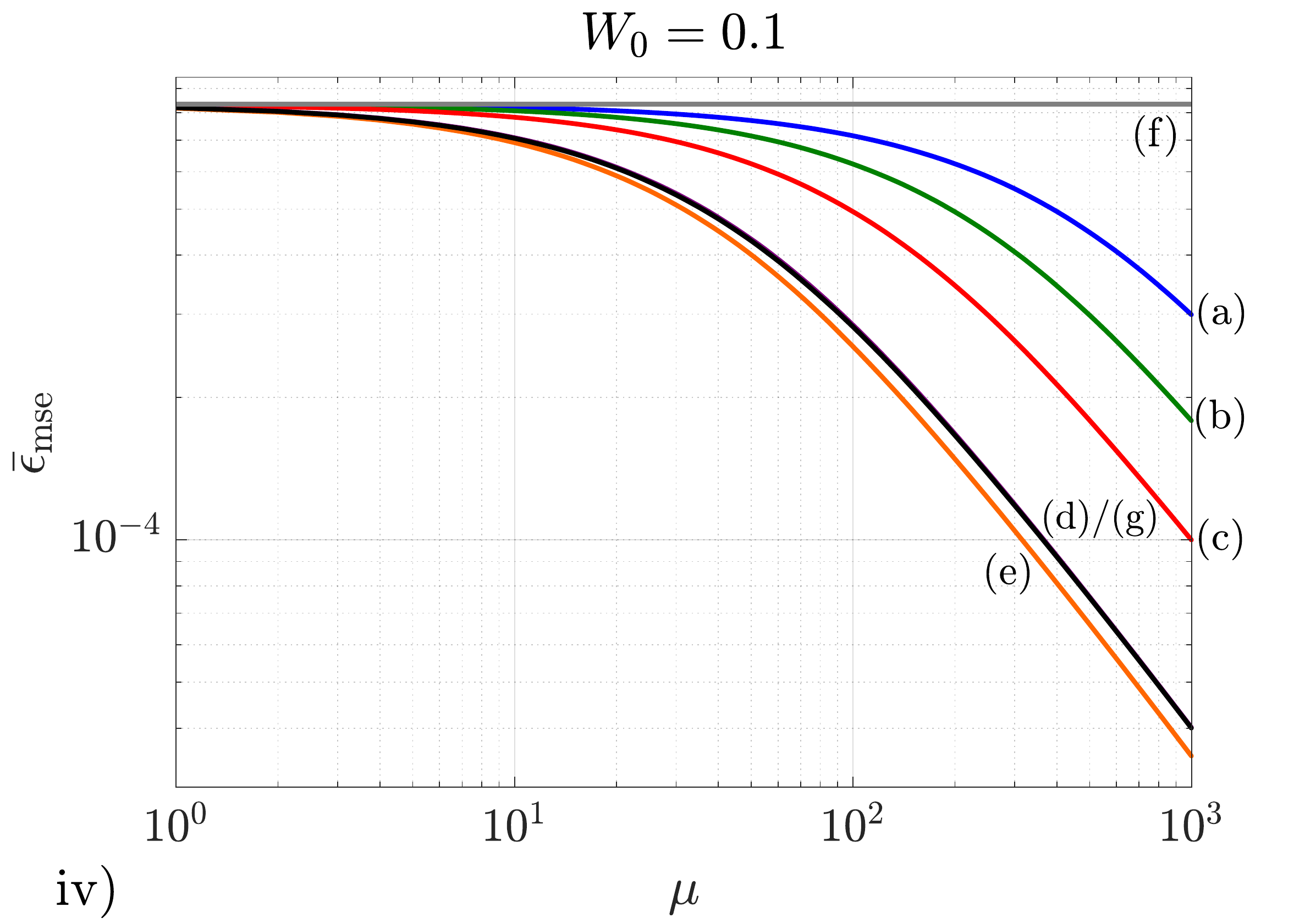}
	\caption{i) Mean square error as a function of the number of repetitions using the optimal single-shot strategy in equation (\ref{singleshot_strategy}) for (a) the coherent state, (b) the NOON state, (c) the twin squeezed vacuum state, (d) the squeezed entangled state, (e) the (optimal) twin squeezed cat state, and (g) the (intermediate) twin squeezed cat state, with mean number of photons $\bar{n}=2$, prior mean $\bar{\theta} = 0$ and prior width $W_0 = \pi/2$, while (f) represents the variance of the prior probability; (ii) repetition of the calculation performed in (i) with prior width $W_0=\pi/3$, (iii) $W_0=\pi/4$, and (iv) $W_0=0.1$. The results in these figures represent the transition from the regime of intermediate prior knowledge and a low number of trials to the local regime of a narrow prior and a large number of measurements.}
\label{prior_figure}
\end{figure*}

Taking the same form of the prior probability given in equation (\ref{prior_probability}), the parameters that we can alter are the prior width $W_0$ and the prior mean $\bar{\theta}$. In section \ref{main_results} we already mentioned that the bounds constructed in figure \ref{bounds_results}.i do not depend on $\bar{\theta}$, leaving $W_0$ as the only free parameter. In principle we should consider the possibility of having both $W_0>\pi/2$, which includes the intermediate and global regimes, and $W_0<\pi/2$, which encompasses the intermediate and local regimes. However, for large values of $W_0$ it is not possible to approximate the periodic error function in equation (\ref{singleshotsinerror}) to the mean square error in equation (\ref{singleshotmse}). For that reason, we will only focus on the transition from the intermediate regime of prior knowledge to the local regime. 

To do this, let us start by calculating the optimal single-shot mean square error in equation (\ref{singleshot_bound}) for all the states with the prior widths $W_0 = \pi/2,~\pi/3,~\pi/4$ and $0.1$. The numerical results are shown in table \ref{prior_effect_summary}. While the best probe in the single-shot regime for $W_0 = \pi/2$ is the twin squeezed vacuum state, the squeezed entangled state becomes the preferable choice when $W_0 = \pi/3$ and $W_0 = \pi/4$, and we need to start with a prior with width $W_0 = 0.1$ in order to recover the twin squeezed cat state as the optimal state. Moreover, the ordering of probes in terms of their performance when $W_0 = 0.1$ is exactly the same as the ordering found in the asymptotic regime, which is also included in the last column of table \ref{prior_effect_summary}. Consequently, we can say that for our schemes the local regime due to a high amount of prior information is achieved when $W_0 \leqslant 0.1$.

An equivalent path to arrive to the same result relies on the approximation
\begin{equation}
\bar{\epsilon}_{\mathrm{mse}}\gtrsim\Delta \theta_p^2\left(1-\Delta\theta_p^2 F_q\right)
\label{bayes_bound_high_prior}
\end{equation}
for the single-shot mean square error employed in \cite{macieszczak2014bayesian, jarzyna2015true}. This relation was found in \cite{macieszczak2014bayesian} assuming a Gaussian prior with a narrow width but, in fact, in appendix \ref{prior_appendix} we show that it also holds for the flat prior introduced in equation (\ref{prior_probability}) if we assume that $W_0 \ll 1$. That the Fisher information appears as the key quantity to determine which scheme has the best performance for a given prior explains why the numerical results in table \ref{prior_effect_summary} for $W_0=0.1$ predict the same order of probes as the approximation $1/(\mu F_q)$ in the asymptotic regime of many repetitions. In both cases, the larger $F_q$, the better the performance. 

It is interesting to observe the similarity between the local regime of prior information for a single shot and the local regime due to a large number of experiments. On the one hand, the best states for $W_0=\pi/2$ and $W_0=0.1$ have intra-mode correlations only, while for $W_0 = \pi/3$ and $W_0 = \pi/4$ the best state presents path entanglement too. On the other hand, figure \ref{prior_figure}.i shows that for $1\leqslant\mu<5$ and $\mu>40$ there is no inter-mode entanglement in the optimal probes, but it appears in the best state for $5<\mu<40$. One way of understanding this similar behaviour is to observe that updating our posterior density via Bayes' theorem after each new trial reduces the uncertainty in a way that is formally similar to making the prior narrower in a sequential way. Nevertheless, both processes are conceptually different. 

Finally, figures \ref{prior_figure}.i - \ref{prior_figure}.iv demonstrate the transition from the intermediate regime of prior knowledge and a low number of trials to a local regime with both high prior information and a large number of repetitions. This process modifies the connection between the number of repetitions and the properties of different probes considerably, as can be seen by the change in the points where the graphs for different states cross each other as the prior width is reduced. As a consequence, establishing a pattern that helps us to understand what probes we need to use for different values of $\mu$ in the regime of limited data becomes more complicated than in the two previous sections. Fortunately, this is not a problem in real experiments because we typically know what our specific prior information is and we can always proceed on a case-by-case basis, but it constitutes an important obstacle to deriving more general results.

\begin{figure*}[t]
\centering
\includegraphics[trim={0.1cm 0.1cm 1.3cm 0.2cm},clip,width=9.1cm]{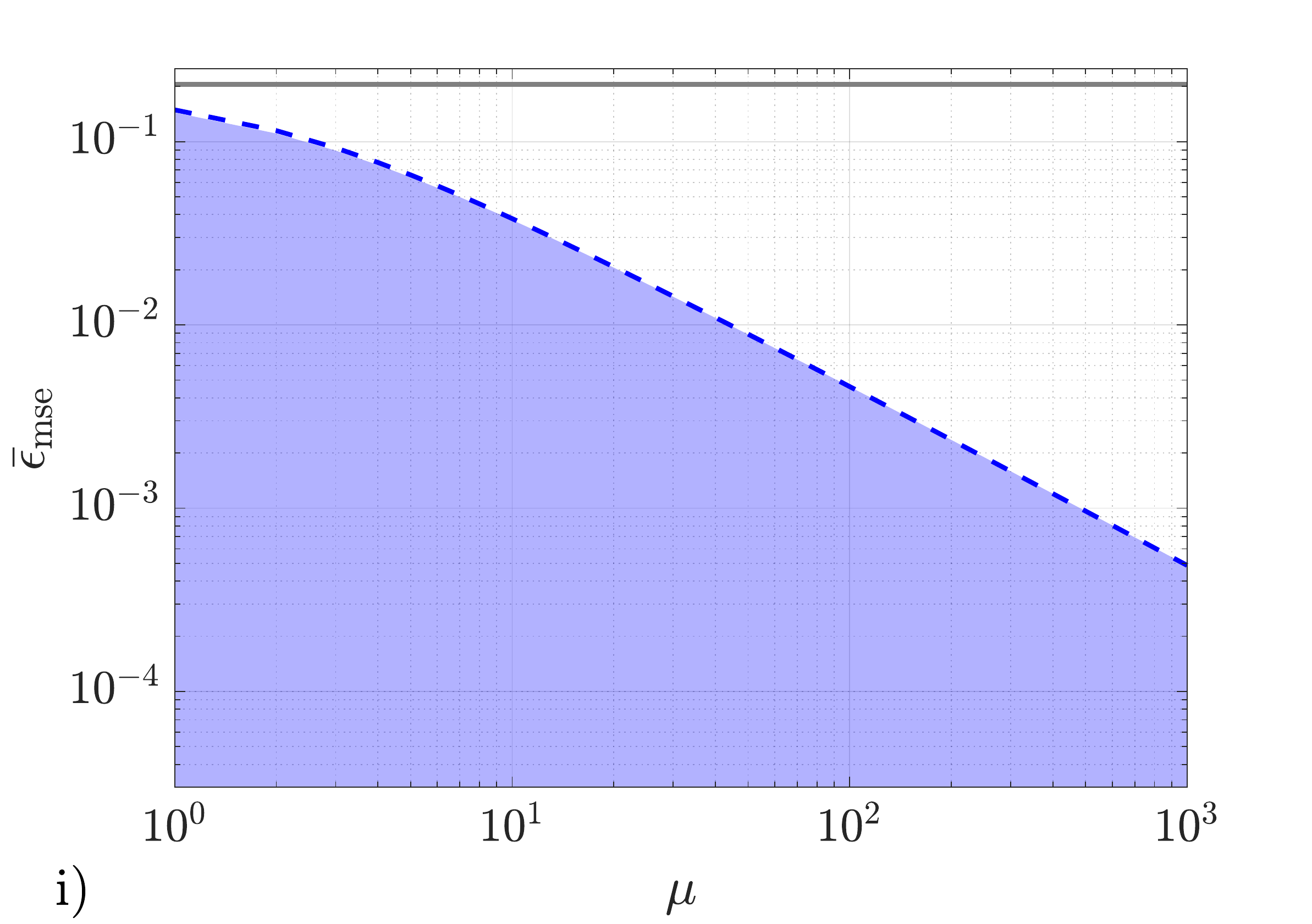}\includegraphics[trim={0.1cm 0.1cm 1.3cm 0.2cm},clip,width=9.1cm]{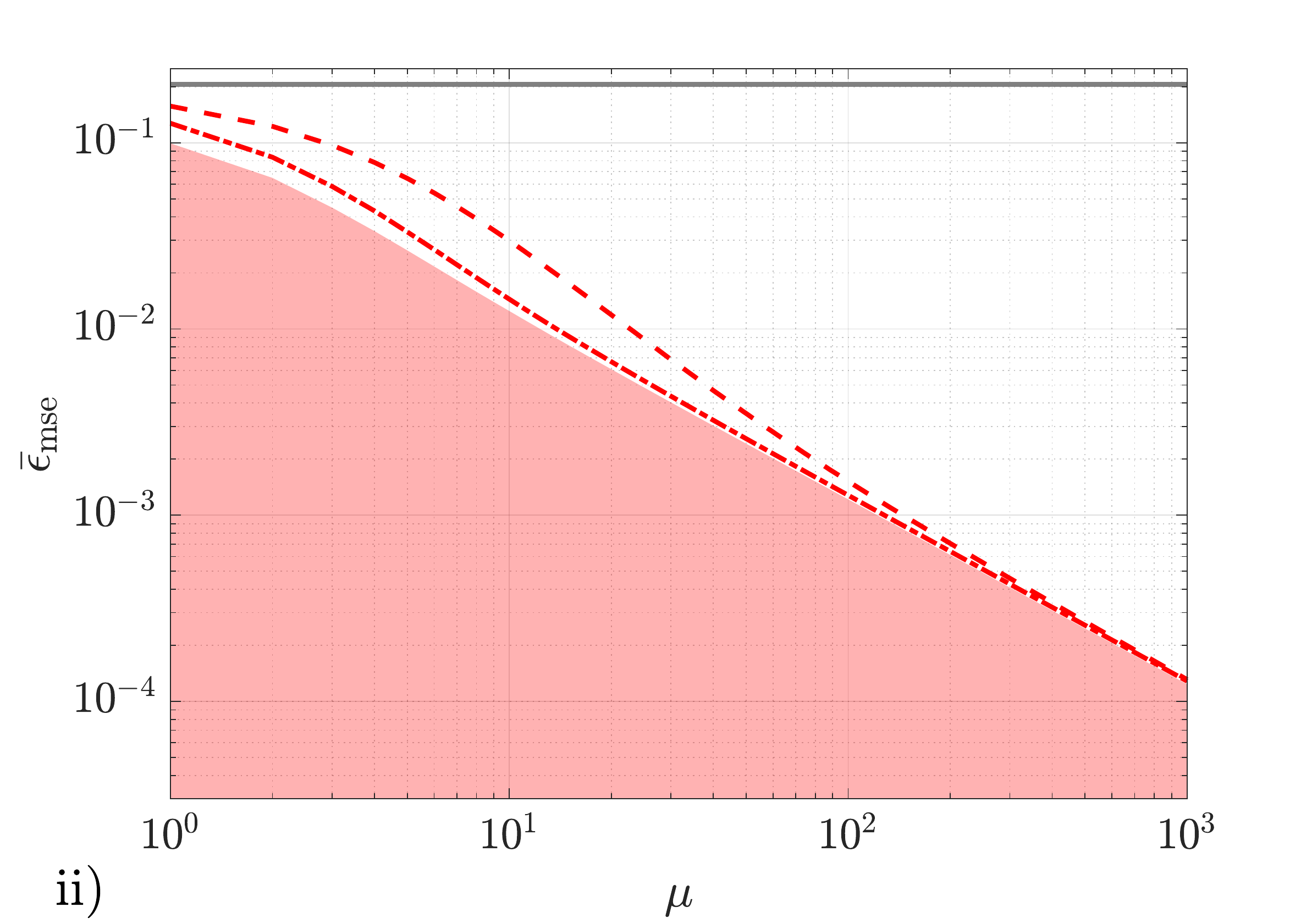}
\includegraphics[trim={0.1cm 0.1cm 1.3cm 0.2cm},clip,width=9.1cm]{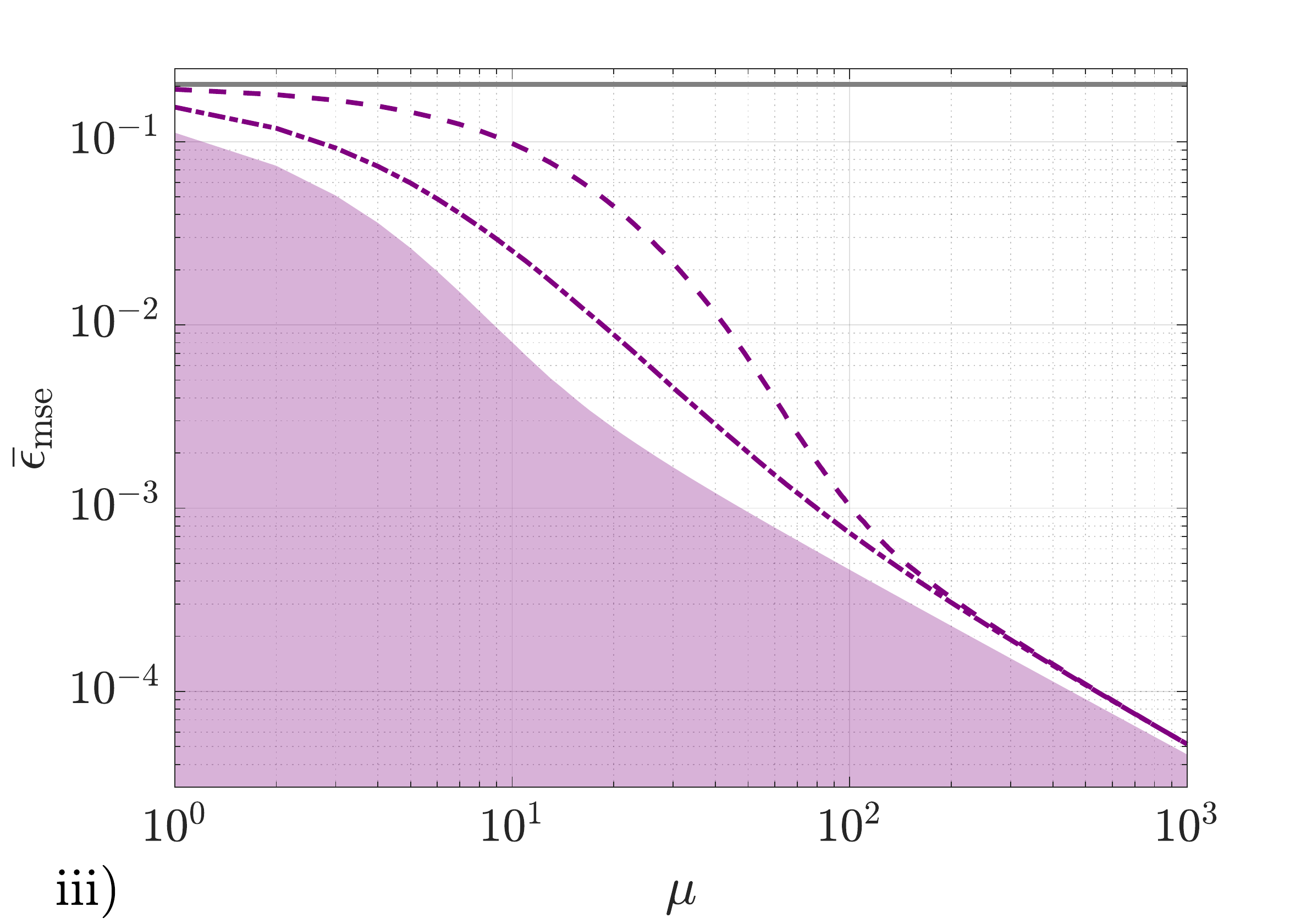}\includegraphics[trim={0.1cm 0.1cm 1.3cm 0.2cm},clip,width=9.1cm]{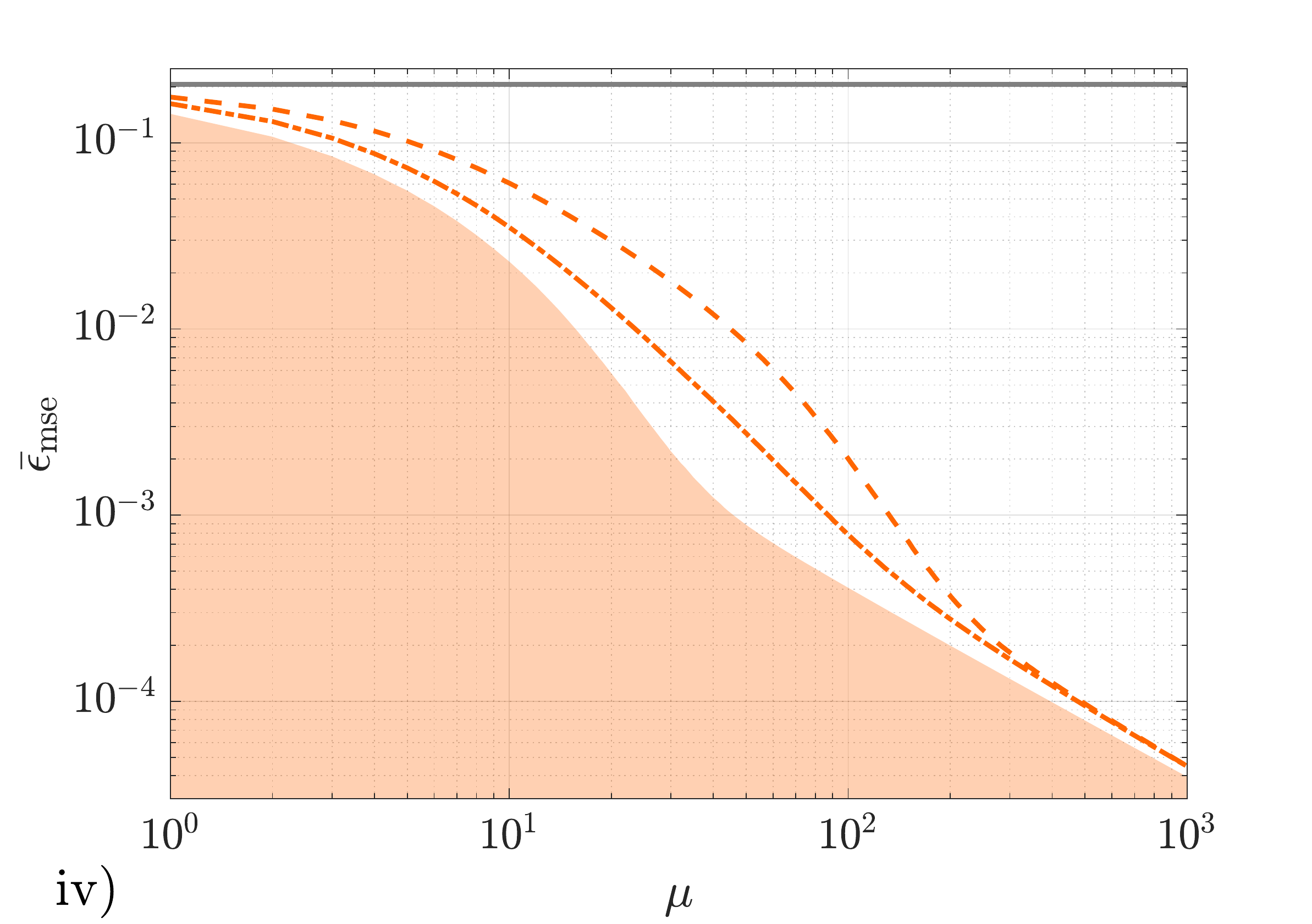}
	\caption{i) Mean square error based on the optimal single-shot strategy (shaded area), error associated with the measurement of energy (dashed line) and prior variance (horizontal solid line) for the coherent state, (ii) the twin squeezed vacuum state, (iii) the squeezed entangled state, and (iv) the twin squeezed cat state, with mean number of photons $\bar{n}=2$, prior mean $\bar{\theta} = 0$ and prior width $W_0 = \pi/2$. Furthermore, the dash-dot graphs in (ii), (iii) and (iv) represents the uncertainty for the measurement of quadratures. The sequences of operations that implement the POVMs that produce these results can be found in table \ref{povm_summary}.}
\label{continuouspovm}
\end{figure*}

\section{Physical measurements}\label{measurements_section}

Until now we have investigated the physical consequences of the bounds constructed following the procedure of section \ref{theory}. Nevertheless, in a real-world situation we also need to be able to generate concrete sequences of operations that can be implemented in the laboratory, study whether they saturate the theoretical bounds and, if they do not, determine how close to the fundamental minimum the associated uncertainty is. Since here we are using a fixed set of probe states, we need only consider sequences for implementing the measurement scheme.

\subsection{Practical states} 

States that can be generated using operations such as squeezing or displacement from the vacuum are generally easier to prepare in the laboratory than the abstract (and possibly entangled) probe states that arise in theoretical optimisations \cite{rafal2015,PaulProctor2016, schafermeier2018} and, as a consequence, there is a intrinsically practical interest in exploring how close to the fundamental bounds this type of state can get. We already know that we can approach the quantum Cram\'{e}r-Rao bound  asymptotically for path-symmetric pure (but otherwise general) states when each individual measurement consists of counting photons after the action of a $50$:$50$ beam splitter \cite{HofmannHolger2009, jesus2017}. For instance, using that POVM it was shown in \cite{jesus2017} that if $W_0 = \pi/2$ and we impose that the relative error in equation (\ref{threshold}) is $\varepsilon_{\tau}=0.05$, then this is true for the twin squeezed vacuum state for $\mu_\tau \geqslant 874$, although surpassing the $0.05$ threshold with the squeezed entangled state requires more than $\mu = 10^3$ repetitions because its convergence is slower.

By using the bounds with $W_0=\pi/2$ and $\bar{\theta} = 0$ in section \ref{main_results} we can now answer this question in the regime of limited data too, both for the previous states and for the coherent and the twin squeezed cat states. As a preliminary step we have reproduced these bounds as shaded areas in figures \ref{continuouspovm}.i - \ref{continuouspovm}.iv for the coherent state, the twin squeezed vacuum state, the squeezed entangled state and the twin squeezed cat state, respectively. In addition, the dashed lines in those figures represent the mean square error associated with the measurement of the energy at each port of the interferometer (i.e., counting photons) after the action of a $50$:$50$ beam splitter. We draw attention to the fact that we have also introduced a known phase shift in the second port of the interferometer before this beam splitter is applied, the complete sequence of operations for each state being presented in table \ref{povm_summary}. The reason behind this choice is that we have found that the uncertainty of this POVM depends on $\bar{\theta}$, and the extra phase shift allows us to achieve the optimal single shot precision when the prior is centred around $\bar{\theta} = 0$, which is our case \footnote{The results in \cite{jesus2017} for $W_0=\pi/2$ were based on a flat prior centred around $\pi/4$ and did not include the extra phase shift discussed in this work as part of the measurement scheme. However, this configuration is numerically equivalent to the one discussed here, and the comparison between both works is thus meaningful for $W_0=\pi/2$.}. This dependence with $\bar{\theta}$ can be seen as a Bayesian analogue of those cases where the standard error propagation formula for a given observable depends on the unknown parameter $\theta$, which is not a problem in practice provided that the experiment is arranged close to an optimal operating point \cite{rafal2015}.

\begin{table*} [t]
{\renewcommand{\arraystretch}{1.45} 
\begin{tabular}{|l|c|c|c|}
\hline
Measurement scheme & Observable & Projectors  & Probes \\
\hline
\hline
\begin{tabular}{@{}l@{}}$50$:$50$ beam splitter \& \\ photon counting (even) \end{tabular} & $N_1 N_2 = \int dk~k \ketbra{k}{k}$, with $N_i = a_i^{\dagger}a_i$ & $\left\lbrace\mathrm{exp}\left(-i\frac{\pi}{4}N_2\right)\mathrm{exp}\left(-i\frac{\pi}{2}J_x\right)\ket{k}\right\rbrace_{k}$ & \begin{tabular}{@{}c@{}} All except \\ coherent state \end{tabular} \\
\hline
\begin{tabular}{@{}l@{}}$50$:$50$ beam splitter \& \\ photon counting (odd) \end{tabular} & $N_1 N_2 = \int dk~ k \ketbra{k}{k}$, with $N_i = a_i^{\dagger}a_i$ & $\left\lbrace\mathrm{exp}\left(-i\frac{\pi}{2}N_2\right)\mathrm{exp}\left(-i\frac{\pi}{2}J_x\right)\ket{k}\right\rbrace_{k}$ & Coherent state \\
\hline
$\pi/8$ - quadratures & \begin{tabular}{@{}c@{}}$X_1 X_2 = \int dq~ q \ketbra{q}{q}$, with \\ $X_i = [ \mathrm{exp} \left( i \frac{\pi}{8} \right) a_i^{\dagger} + \mathrm{exp}\left(-i\frac{\pi}{8}\right)a_i]/\sqrt{2}$\end{tabular}  & $\left\lbrace\mathrm{exp}\left(i\frac{\pi}{4}N_1\right)\mathrm{exp}\left(-i\frac{\pi}{2}J_x\right)\ket{q}\right\rbrace_{q}$ &  \begin{tabular}{@{}c@{}} All except \\ coherent state \end{tabular} \\
\hline
\begin{tabular}{@{}l@{}}Undoing preparation \\ \& photon counting\end{tabular} & $N_1 N_2 = \int dk~ k \ketbra{k}{k}$ & $\left\lbrace\mathrm{exp}\left(i \pi J_z \right)\mathrm{exp}\left(i\frac{\pi}{2}J_x\right)D_1^{\dagger}\left(\alpha\right)\ket{k}\right\rbrace_{k}$ & Coherent state \\
\hline
Parity measurement & $\Pi_1 \Pi_2 = \int dp~ p \ketbra{p}{p}$, with $\Pi_i = (-1)^{a_i^{\dagger}a_i}$ & $\left\lbrace\mathrm{exp}\left(-i\frac{\pi}{4}N_2\right)\mathrm{exp}\left(-i\frac{\pi}{2}J_x\right)\ket{p}\right\rbrace_{p}$ & NOON state \\ 
\hline
\end{tabular}}
\caption{Sequences of quantum operations needed to implement the practical measurements discussed in section \ref{measurements_section}, whose uncertainty is represented in figures \ref{continuouspovm} and \ref{noonpovm}. Note that the observable column indicates the physical quantity that is being measured, and that the different combinations of phase shifts that appear in the third column have been chosen such that the schemes are optimal when the prior is centred around $\bar{\theta} = 0$ and $\bar{n} = 2$.}
\label{povm_summary}
\end{table*}

To start our discussion of the low trial number regime with this POVM, we first observe that, according to figure \ref{continuouspovm}.i, measuring energy with coherent states produces an uncertainty that is already very close to the associated bound for a low value of $\mu$. More concretely, the bound and the measurement error only differ in their second and third significant figures, as can be directly verified from the numerical values in table \ref{practical_povm_summary} that we provide in appendix \ref{numcal} for the first ten shots of every scheme based on indefinite photon number strategies. Moreover, this can be further improved if instead we undo the preparation of the probe state before counting photons, that is, by reversing the $50$:$50$ beam splitter and the displacement from the vacuum operations that generated the coherent state in the first place. The extra known difference of phases showed in table \ref{povm_summary} is also needed for the case with $\bar{\theta}=0$ that we are considering. Nonetheless, taking into account the fact that both schemes produce an uncertainty whose first significant figure is that of the optimum (see table \ref{practical_povm_summary}), we conclude that, for most practical purposes, they are equally useful and optimal given any number of repetitions. 

The situation is very different when we consider the other three states in figures \ref{continuouspovm}.ii - \ref{continuouspovm}.iv, where the uncertainty of the energy measurement is now notably higher than each bound in the regime of limited data, the distance between the graphs of the measurement and those of the bounds being larger for a few repetitions than for a single shot. This measurement is particularly detrimental for the strategy based on the squeezed entangled state, since its error is very close to the prior variance (horizontal line in \ref{continuouspovm}.iii) when $\mu \sim 1$ and this indicates that almost no information is being gained there. Additionally, we can observe that the twin squeezed cat state in figure \ref{continuouspovm}.ii presents a slow convergence to the asymptotic Cram\'{e}r-Rao bound when we use this POVM, compared with the twin squeezed vacuum probe state in figure \ref{continuouspovm}.ii or the coherent state in in figure \ref{continuouspovm}.i. Note that this is the same problem found in \cite{jesus2017} for the squeezed entangled state, which is also reproduced in our calculations here.

These results show that counting photons is not the best strategy to be followed when $\mu$ is low and the probes have been prepared in states with a large Fisher information such as the ones considered here, and this motivates the search for other practical alternatives. More concretely, instead of projecting onto the energy basis, we can consider the measurement of a different physical quantity. The dash-dot lines in figures \ref{continuouspovm}.ii - \ref{continuouspovm}.iv show the results where we have projected onto the eigenvectors of the observable $X_1\otimes X_2$,
\begin{equation}
X_i =\frac{1}{\sqrt{2}}\left(a_i^\dagger \mathrm{e}^{-i\pi/8} + a_i \mathrm{e}^{i\pi/8}\right)
\label{quadrature}
\end{equation}
being a quadrature rotated by $\pi/8$ for the $i$-th mode \cite{barnett2002}, after having introduced the phase shift $\mathrm{exp}(i \frac{\pi}{4}a_1^\dagger a_1)$ and having applied a $50$:$50$ beam splitter (see table \ref{povm_summary}) \footnote{Note that the eigenstates of the quadrature operator in equation (\ref{quadrature}) cannot be normalised \cite{barnett2002}, and that the eigenvectors mentioned in the main text refer to the numerical approximation associated with the truncated Hilbert state that we are employing here.}. The error of this scheme also depends on $\bar{\theta}$.

By comparing the energy and quadrature measurements in figures \ref{continuouspovm}.ii - \ref{continuouspovm}.iv we see that the graphs based on the latter POVM are substantially closer to the bounds than those for the former measurement when the experiment is operating in the regime of limited data. In other words, we have found a physical measurement that improves over the results based on measuring the energy for the practical states under consideration and a low number of trials. Interestingly, the dash-dot lines still converge to the fundamental asymptotic bound, and this implies that in the asymptotic regime both schemes are, nevertheless, equivalent in practice and optimal.   

Although these results extend, generalise and clarify the findings of \cite{jesus2017}, figures \ref{continuouspovm}.ii - \ref{continuouspovm}.iv also show that it could still be possible to find other physical schemes with a better precision when $\mu$ is low, with a faster rate of convergence to the asymptotic minimum or even saturating the bound for any $\mu$. These are some of the key questions that should be addressed for further progress in the design of experimentally feasible protocols that operate both in and out of the regime of limited data. 

\subsection{Optimality of NOON states}

\begin{figure*}[t]
\centering
\includegraphics[trim={0.1cm 0.1cm 1.3cm 0.5cm},clip,width=9.1cm]{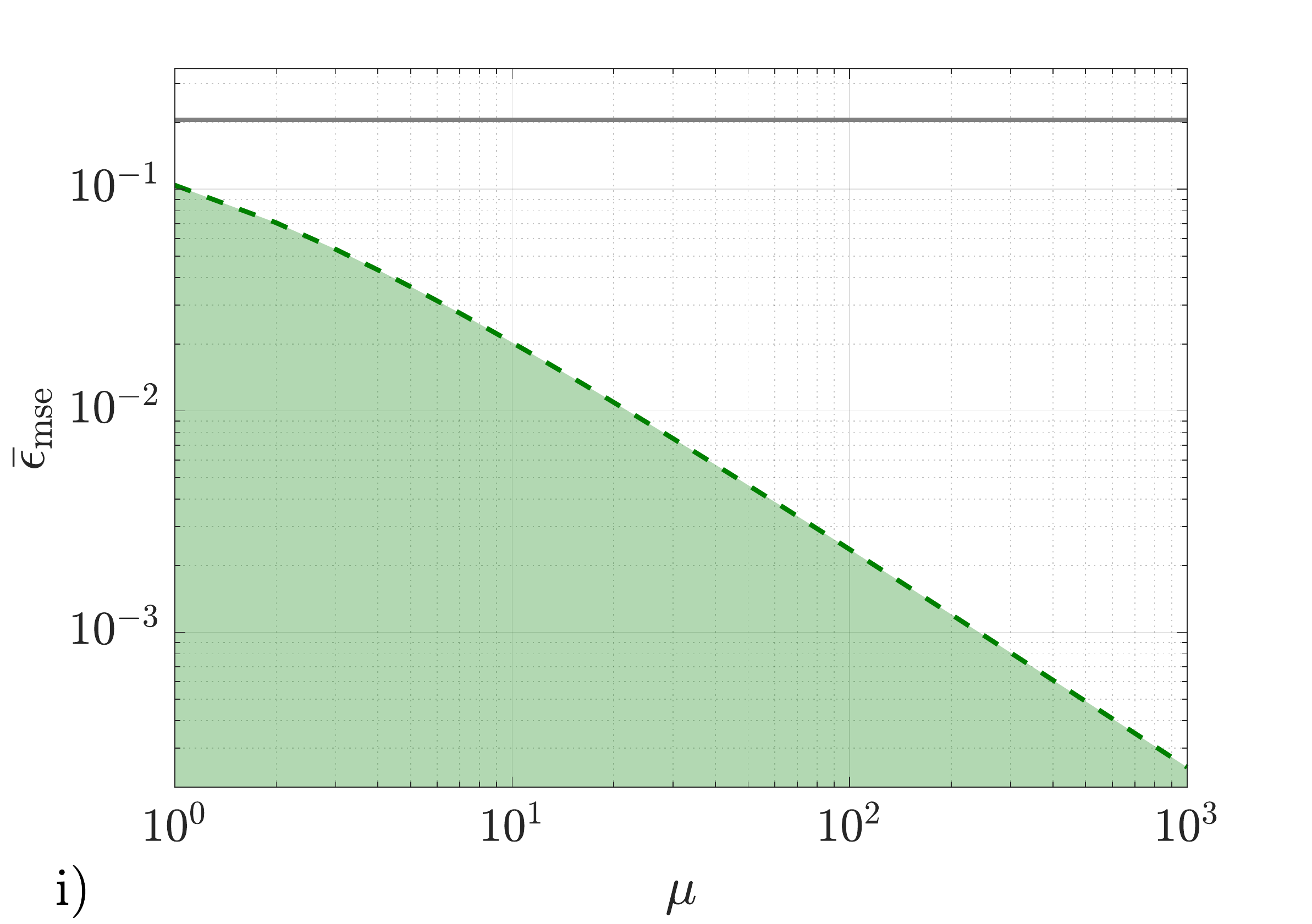}\includegraphics[trim={0.1cm 0.1cm 1.3cm 0.5cm},clip,width=9.1cm]{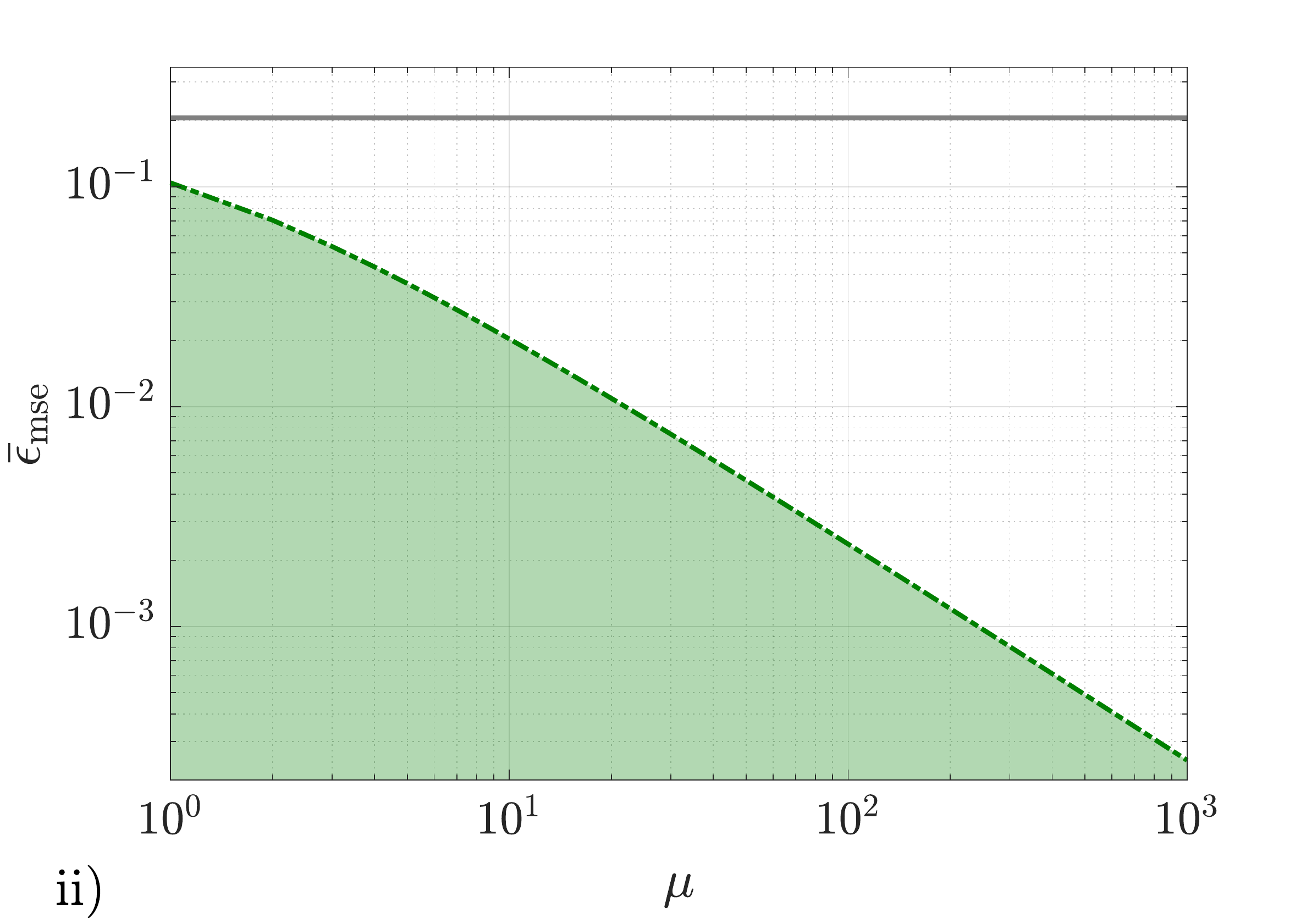}
\includegraphics[trim={0.1cm 0.1cm 1.3cm 0.5cm},clip,width=9.1cm]{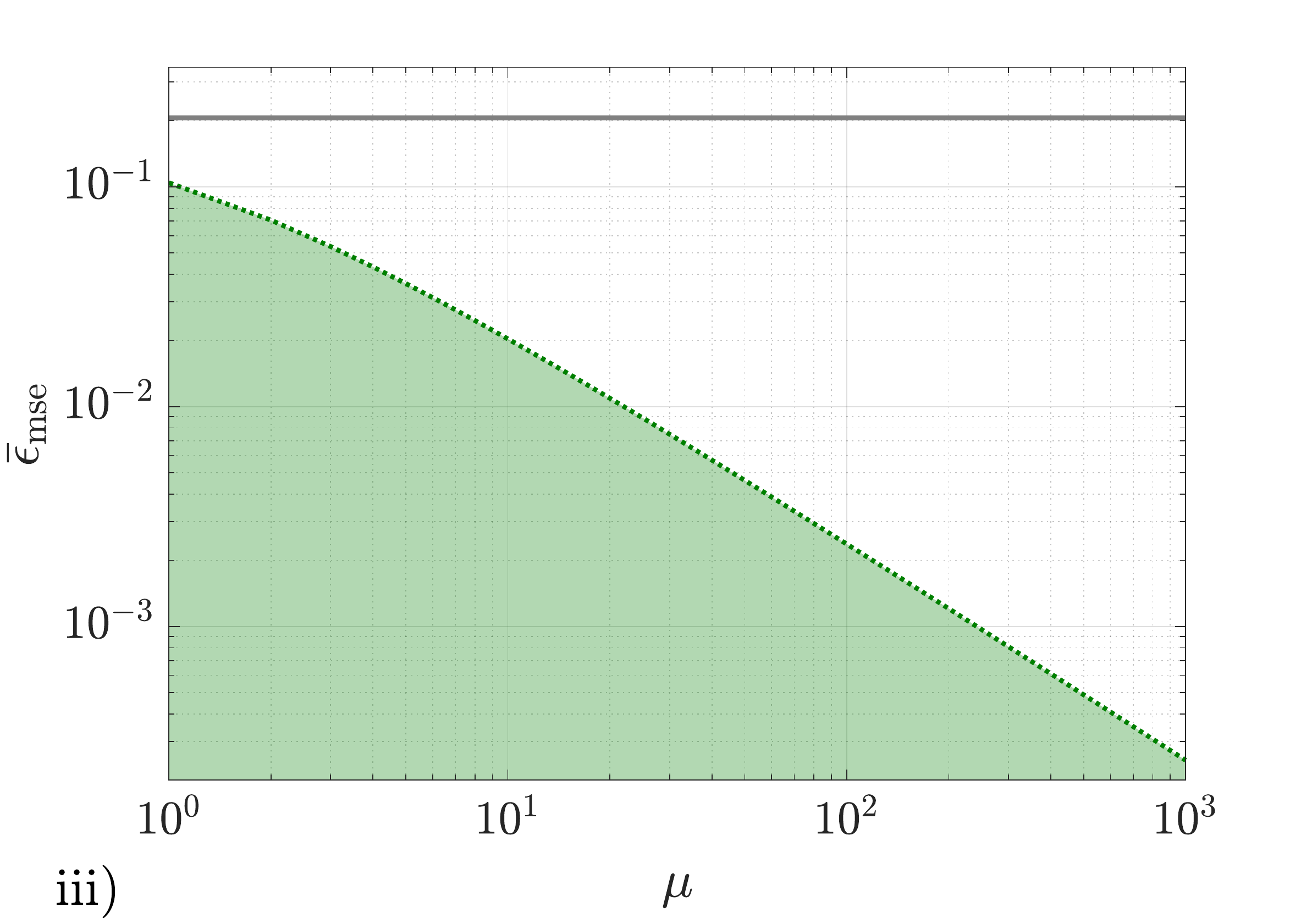}\includegraphics[trim={0.1cm 0.1cm 1.3cm 0.5cm},clip,width=9.1cm]{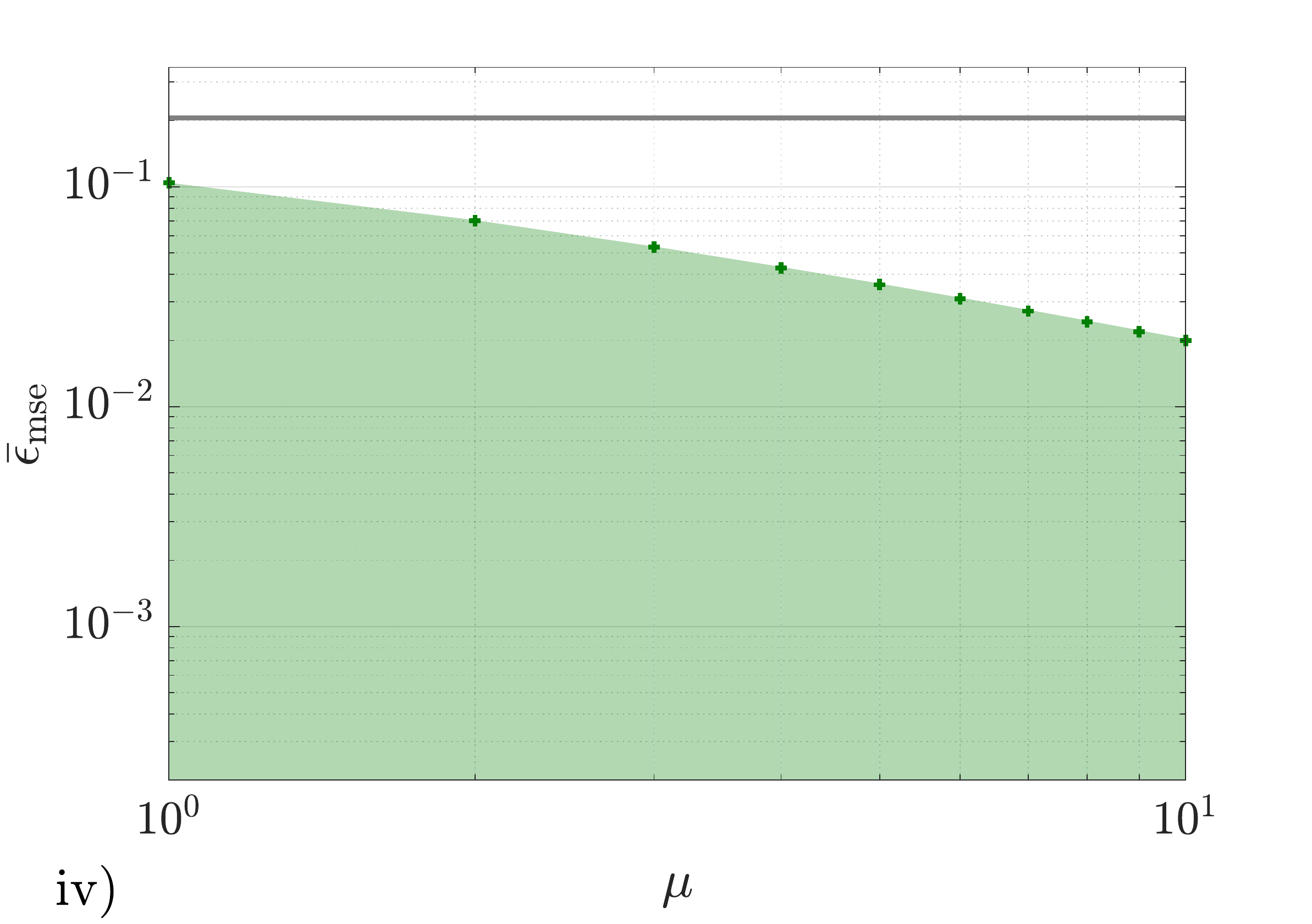}
	\caption{Mean square error based on the optimal single-shot strategy (shaded area), prior variance (horizontal solid line) and error associated with (i) the measurement of energy (dashed line), (ii) the measurement of quadratures (dash-dot line), (iii) parity measurements (dotted line), and (iv) the optimal collective measurement on $\mu$ copies of the probe (plus signs), for a NOON probe state with $\bar{n}=2$, $\bar{\theta} = 0$ and $W_0 = \pi/2$.}
\label{noonpovm}
\end{figure*}

The fact that NOON states are conceptually simple makes them an excellent tool to understand metrology protocols, which is why we have chosen to study them separately. They emerge as the optimal probe when we maximise the Fisher information over the definite photon number states \cite{demkowicz2011, jarzyna2016thesis}, and while they are unsuitable for a global estimation due to the multi-peak structure associated with the posterior probability functions that they generate \cite{jesus2017, jarzyna2016thesis, kolodynski2014}, and they require that the scaling of the prior variance is already $\sim 1/\bar{n}^2$ in order to achieve the same scaling that the Cram\'{e}r-Rao bound predicts \cite{berry2012, hall2012}, the results in \cite{jesus2017} showed that they can still be useful to a certain extent in the intermediate regime of prior knowledge and limited data when the number of photons is low and the POVM is based on measuring the energy at each port. In addition, this moderate usefulness also holds for the repetition of the single-shot optimal strategy according to our previous results in figure \ref{bounds_results}.i, since the NOON state performs better than the twin squeezed cat state for $1\leqslant\mu\leqslant10$. By studying the performance of this probe for different physical measurements with respect to the non-asymptotic bound we will see that NOON states are also optimal in another sense. 

First we consider the two measurement schemes that we described in the previous section, that is, counting photons and measuring rotated quadratures after the introduction of some phase shifts that are indicated in table \ref{povm_summary}, and after the action of a $50$:$50$ beam splitter. The mean square errors generated by them for the NOON state, which are represented in figures \ref{noonpovm}.i and \ref{noonpovm}.ii, respectively, display a perfect agreement with the bounds for any number of repetitions. This can be further verified by observing that the uncertainties for the first ten shots provided in table \ref{noon_povm_summary} of appendix \ref{numcal} are virtually identical.

Similarly, a parity measurement based on the projectors of the observable $\Pi_1\otimes \Pi_2 =  (-1)^{a_1^{\dagger}a_1}\otimes (-1)^{a_2^{\dagger}a_2}$ \cite{gerry2010, chiruvelli2011}, and performed after introducing an extra phase shift and the action of a beam splitter (see table \ref{povm_summary}), also saturates the bound for all $\mu$, as it can be observed in figure \ref{noonpovm}.iii. This is consistent with the fact that the information about the phase is actually contained in the parity of the number of photons \cite{kolodynski2014, gerry2010, chiruvelli2011}. Interestingly, we have verified that counting photons and checking the parity at each port produces the same non-asymptotic results for the indefinite photon number states too. 

That different physical schemes are able to saturate the same quantum bound can be explained by recalling that the optimal quantum estimator $S$ is only defined on the support of $\rho$ (see \cite{Note53} and appendix \ref{numcal}). In particular, for NOON states $\rho$ can be represented by a non-singular ($2 \times 2$) matrix in the number basis (see appendix \ref{noon_analytical}), which is only a part of the full Hilbert space including all the sectors with any number of photons. As a consequence, any measurement scheme that coincide with the projectors $\ket{s_1}$ and $\ket{s_2}$ given in equation (\ref{noon_projectors}) in the part of the Hilbert space that corresponds to the support of $\rho$ is going to be optimal, independently of the particular form of the POVM elements. 

Furthermore, the same intuition can be used to understand why it is more difficult to saturate the bounds for indefinite photon number states in the non-asymptotic regime. For these states there is a non-zero probability of detecting any number of photons at each port of the interferometer, which implies that the optimal quantum estimator $S$ can be constrained in all the sectors of the Hilbert space, and these constraints need to be fully satisfied to saturate the single-shot bound. However, as we accumulate more data we start to approach the quantum Cram\'{e}r-Rao bound, which is based on the equation $L(\theta) \rho(\theta) + \rho(\theta) L(\theta) = 2 \partial \rho(\theta)/\partial\theta$, and this equation only has a unique solution on the support of $\rho(\theta)$ \cite{genoni2008}, which in our case is simply a pure state. That is, finding physical measurements that saturate the asymptotic bounds is generally less demanding and, in fact, the errors of the physical measurements in figures \ref{continuouspovm}.i - \ref{continuouspovm}.iv converge to the fundamental bound. 

This state of affairs gives raise to an interesting situation. The Bayesian bounds in figure \ref{bounds_results}.i show that, in principle, the NOON state is not the best option among the probes that we are examining for any number of repetitions. In spite of this fact, if we compare the uncertainty associated with counting photons after undoing the preparation of a coherent state, the measurement of quadratures for the states based on the squeezing operator,  and any of the physical measurement previously discussed for the NOON state, then it can be shown that, in this case, the NOON state is the best probe when $1 \leqslant \mu \leqslant 3$. In particular, this conclusion can be extracted by inspection from tables \ref{practical_povm_summary} and \ref{noon_povm_summary} in appendix \ref{numcal}. This analysis highlights the importance of studying the possibility of saturating the theoretical bounds using realistic implementations in a particularly transparent way.

On the other hand, the mathematical simplicity of NOON states allows us to go one step further and study collective measurements \cite{jarzyna2015true, jarzyna2016thesis}. Until now this work has been based on preparing some probe, implementing its optimal strategy in equation (\ref{singleshot_strategy}) and repeating this procedure $\mu$ times. However, a more general possibility is to prepare $\mu$ identical states and perform a single measurement on all of them at once. If we upgrade the optimal single-shot bound in equation (\ref{singleshot_bound}) to cover the collective case we find that
\begin{equation}
\bar{\epsilon}_\mathrm{mse} \geqslant \int d\theta p(\theta) \theta^2 - \mathrm{Tr}\left(\bar{\rho}_\mu S_\mu\right),
\label{singleshot_collective}
\end{equation} 
where now $S_\mu$ is given by $S_\mu \rho_\mu + \rho_\mu S_\mu = 2 \bar{\rho}_\mu$ with 
\begin{equation}
\rho_\mu = \int d\theta p(\theta) \smash[b]{ \underbrace{\rho(\theta) \otimes \cdots \otimes \rho(\theta)\,}_\text{$\mu$ times}}
\label{zerothmoment_collective}
\end{equation}
and
\begin{equation}
\bar{\rho}_\mu = \int d\theta p(\theta) \smash[b]{ \underbrace{\rho(\theta) \otimes \cdots \otimes \rho(\theta)\,}_\text{$\mu$ times}}\theta.
\label{firstmoment_collective}
\end{equation}

A calculation scheme for equation (\ref{singleshot_collective}) is proposed in appendix \ref{numcal_noon}, and its application to the NOON state for $1\leqslant\mu\leqslant10$ results in the graph of figure \ref{noonpovm}.iv, which coincides with the bound generated by repeating the optimal strategy for a single probe. Numerically, this agreement occurs at least for the first significant figure, as it can be verified in table \ref{noon_povm_summary} available in appendix \ref{numcal}. We conclude thus that collective measurements do not provide a better performance than the practical measurements previously studied when we are working in the low-$\mu$ regime, each probe is prepared in a NOON state with $\bar{n} = 2$ and the prior width is $W_0=\pi/2$.

In summary, we have shown that there are measurement schemes that can saturate the bound for the NOON state for all $\mu$ simultaneously. Consequently, NOON states do not only have a special status in the local regime, but also in the regime of limited data and moderate prior knowledge \footnote{In \cite{jarzyna2016thesis} it is argued that NOON states emerge as the optimal probe with a definite number of photons in the limit where the prior information dominates, something that is shown in \cite{demkowicz2011}, and it is concluded that, for that reason, using NOON states is almost useless in a practical Bayesian context. Although it is true that NOON states are limited due to the ambiguity that they introduce in the estimation and, more importantly, because of the difficulties to use them in real experiments \cite{schafermeier2018}, we draw attention to the fact that the regime where the prior knowledge may play a substantial role is relevant and useful whenever we need to make inferences from a practical scenario where only a low number of experiments can be performed.}. This can be explained by noticing that the optimal projectors for a single shot in equation (\ref{noon_projectors}) are the same that the projectors predicted by the symmetric logarithmic derivative that define the quantum Fisher information \cite{kolodynski2014}. While this probe state is fragile and difficult to prepare in more realistic scenarios \cite{schafermeier2018}, these results are still interesting from a fundamental perspective, and they have helped us to understand the problems associated with saturating the bounds of more practical states that we need to overcome in the future. 

\section{Conclusions}\label{conclusions}

We have developed a method to study the performance of metrology protocols that operate in the regime of limited data and moderate prior knowledge. More concretely, we have proposed to use the strategy that is optimal after minimising the single-shot mean square error over all the possible POVMs in a sequence of $\mu$ repeated experiments. Given a state, a Hamiltonian and a prior probability, we have seen that the bounds that arise from this technique are optimal for the first shot by construction, and that they also start to converge to the quantum Cram\'{e}r-Rao bound when $\mu\sim 10^2$. In addition, we have argued that they can be saturated using measurements that are equivalent to the projectors of the optimal quantum estimator $S$ for each repetition, and that this strategy is optimal for those experiments based on identical and independent trials where adaptive techniques or more general measurements are excluded. 

The usefulness of this method in the context of quantum metrology has been demonstrated through the analysis of a Mach-Zehnder interferometer, and we have focused our study on three indefinite photon number states that have been proposed in the literature due to their large Fisher information: the twin squeezed vacuum state, the squeezed entangled state and the twin squeezed cat state. We have found that the twin squeezed vacuum state is the best option when $1\leqslant\mu <5$, $W_0=\pi/2$, and for $\mu = 1$, $W_0=\pi/3$; that the squeezed entangled state is the preferred choice if $5<\mu <40$, $W_0=\pi/2$
and when $\mu = 1$, $W_0=\pi/3$ or $W_0=\pi/4$; and that the twin squeezed cat state recovers its status of best probe due to its largest Fisher information when $\mu > 40$, $W_0=\pi/2$ and $\mu=1$, $W_0 = 0.1$. To the best of our knowledge, a fully Bayesian analysis in the terms explored in this work had not been done before for these probes. 

Using the twin squeezed cat state as a family of probes whose parameters can be modified for given mean number of photons and prior width, we have provided evidence that suggests that increasing the amount of intra-mode correlations, that is, the correlations within each arm of the interferometer, could be detrimental when the number of repetitions is low, which contrasts with the fact that the same type of correlations are actually beneficial in the asymptotic regime. Moreover, we have shown that using a state with less intra-mode correlations and a certain amount of path entanglement such as the squeezed entangled state appears to help to enhance the precision in the non-asymptotic regime without damaging the asymptotic performance in a dramatic way. Therefore, we conjecture that there might exist a more general relationship between the number of trials and the amount of  intra-mode and inter-mode correlations that could indicate how to reduce the uncertainty of the protocols in the regime of limited data.

It has been shown that, for a low number of trials, the usual strategy of counting photons after the action of a beam splitter is optimal for most practical purposes when the probe is prepared in a coherent state, although it does not saturate the non-asymptotic bounds for the other indefinite photon number states. However, we have found that in the latter case the situation can be improved if instead we measure quadratures rotated by $\pi/8$, since this scheme is closer to our bounds for low values of $\mu$. This result is particularly relevant because states prepared with operations such as squeezing or displacement from the vacuum and quadratures measurements are easier to implement in real-world situations. In addition, our calculations indicate that counting photons, measuring quadratures and implementing parity measurements are optimal strategies for any number of repetitions if the probe is in a NOON state, and that collective measurements on the first ten copies of this probe do not provide an advantage over the schemes based on identical and independent experiments.

It is important to note that in this work we have not considered what happens in the presence of noise because our aim was to identify the novel effects that emerge directly from having a low number of trials without the interference of other features, which justifies our focus on ideal schemes. However, a comprehensive study of the effect of noise when the available data is limited is also crucial to model realistic scenarios. Although we leave this analysis for future research, in appendix \ref{loss} we provide an initial test to demonstrate that our method can be also applied to a scheme where photon losses are present, finding that the qualitative behaviour of our results does not seem to change substantially for a reasonable amount of loss.

We believe that these results constitute an important advance towards the creation of a practical and useful methodology that will help us to design optimal metrology experiments taking the finite number of trials into account, and that they could play a crucial role in the design of realistic inference schemes once this method is combined with other features such as the presence of noise, larger numbers of photons, adaptive techniques, state engineering algorithms or multi-parameter systems.

\section*{Acknowledgements}

We acknowledge helpful discussions with Paul Knott, Simon Haine, Alfredo Luis, George Knee, Dominic Branford and Pieter Kok. We also thank Michael Hall and Stefano Pirandola for useful suggestions. This work was funded by the South East Physics Network (SEPnet) and the United Kingdom EPSRC through the Quantum Technology Hub: Networked Quantum Information Technology (grant reference EP/M013243/1).


\bibliographystyle{apsrev4-1}

\bibliography{submission_references_21012019}

\appendix

\section{How large $W_0$ can be such that the use of a quadratic error is justified?}\label{prior_sinapprox_appendix}

The approximation $4~\mathrm{sin}^2\lbrace\left[g(\boldsymbol{n})-\theta\right]/2 \rbrace \approx [g(\boldsymbol{n})-\theta]^2$ relies on the quantity $|g(\boldsymbol{n})-\theta|/2$ being small. Moreover, if $W_0$ is the width of the phase domain, then $|g(\boldsymbol{n})-\theta|/2 \leqslant W_0/2$ within one period. The minimimum requirement that is natural to impose is that the variable for which the Taylor expansion is calculated (i.e., $|g(\boldsymbol{n})-\theta|/2$) is slightly smaller than $1$ at most, which is always the case if the width of our experiment satisfies that $W_0 \lesssim 2$. In principle this would still be a crude approximation if we were interested in the sine function itself. However, the sine error is then integrated over all the possible values for $\theta$ and $\boldsymbol{n} = (n_1, \dots , n_\mu)$. This implies that $|g(\boldsymbol{n})-\theta|/2 \sim 2$ when $W_0\sim2$ only for a few combinations of values, and the weight of those cases will decrease as the joint probability $p(\theta,\boldsymbol{n})$ accumulates more data. We conclude then that $W_0 \lesssim 2$ is a reasonable estimation for the range of validity of the mean square error in a problem with a periodic parameter. Note that this condition has the same order of magnitude than the estimation found in \cite{friis2017}, where the authors argued that the width of their Gaussian prior had to be $\pi/2$ or less, and it is a better estimation than the one obtained in \cite{jesus2017}. 

\begin{figure}[t]
\centering
\includegraphics[trim={0.2cm 0.1cm 1.3cm 0.5cm},clip,width=9cm]{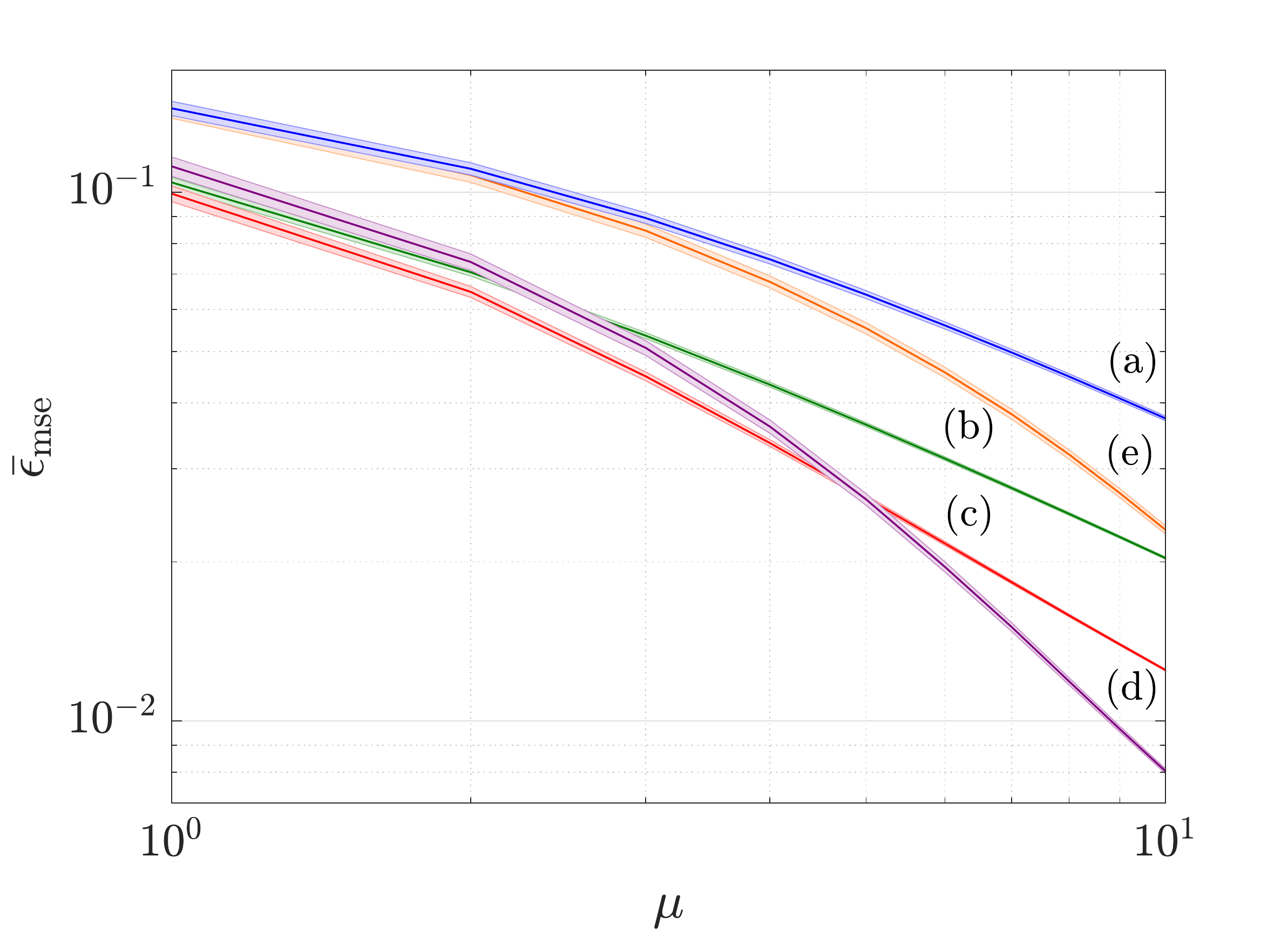}
	\caption{Mean square error based on the optimal single-shot strategy (solid line) and bounds for the approximation error after having expanded the sine error up to second order (shaded area) for (a) the coherent state, (b) the NOON state, (c) the twin squeezed vacuum state, (d) the squeezed entangled state, and (e) the twin squeezed cat state, with $\bar{n} = 2$, $\bar{\theta}=0$ and $W_0 = \pi/2$. This figure shows that the mean square error is a suitable approximation for the mean sine error when we are in the regime of moderate prior knowledge.}
\label{sinapprox_plot}
\end{figure}

According to the previous discussion, only the calculation of the first few shots could be potentially misleading if we use the mean square error. To show that this is not the case for the schemes analysed in the main text, let us estimate explicitly the error of the Taylor expansion. First, using Taylor's theorem we have that $\mathrm{sin}^2(x) = x^2-x^4 \mathrm{cos}(2\varepsilon)/3$, where $\varepsilon \in [0,x]$  \cite{mathematics2004}. The first term is the approximation that we want to use, while the second term represents the error of this approximation. Using the fact that the cosine is bounded between $-1$ and $1$, the Taylor error can be estimated with 
\begin{equation}
\Delta \bar{\epsilon} = \frac{1}{12}\int d\theta d\boldsymbol{n}~p(\theta) p(\boldsymbol{n}|\theta) \left[g(\boldsymbol{n})-\theta\right]^4,
\label{taylorerror}
\end{equation}
and knowing that the optimal phase estimator is the average of the posterior probability $p(\theta|\boldsymbol{n}) \propto p(\theta)p(\boldsymbol{n}|\theta)$, we can rewrite equation (\ref{taylorerror}) as
\begin{equation}
\Delta \bar{\epsilon} = \frac{1}{12}\int d\theta' p(\theta') \int d\boldsymbol{n} p(\boldsymbol{n}|\theta') \Delta \bar{\epsilon}(\boldsymbol{n}),
\end{equation}
where
\begin{equation}
\Delta \bar{\epsilon}(\boldsymbol{n}) = \left\langle \theta^4 \right\rangle- 4 \left\langle \theta \right\rangle \left\langle \theta^3 \right\rangle + 6\left\langle \theta \right\rangle^2 \left\langle \theta^2 \right\rangle - 3\left\langle \theta \right\rangle^4
\end{equation}
and we have used the notation $\left\langle \Box \right\rangle = \int d\theta p(\theta|\boldsymbol{n}) \Box$. This is precisely the three-step decomposition introduced in \cite{jesus2017} to obtain the mean square error and, as such, we can compute $\Delta \bar{\epsilon}$ numerically in the same way.

\begin{table*} [t]
{\renewcommand{\arraystretch}{1.2} 
\begin{tabular}{|c|c|c|c|c|c|}
\hline
\multicolumn{6}{|c|}{$\bar{\epsilon}_{\mathrm{mse}}\left(\mu = 1\right)$, $\dots$, $\bar{\epsilon}_{\mathrm{mse}}\left(\mu = 10 \right)$} \\
\hline
\hline
\multicolumn{3}{|c|}{Coherent state} & \multicolumn{3}{c|}{Twin squeezed vacuum state} \\
\hline
\begin{tabular}{@{}c@{}}Single-shot \\ measurement\end{tabular} & \begin{tabular}{@{}c@{}}$50$:$50$ beam splitter \\ \& photon counting\end{tabular} & \begin{tabular}{@{}c@{}}Undoing preparation \\ \& photon counting\end{tabular} & \begin{tabular}{@{}c@{}}Single-shot \\ measurement\end{tabular} & \begin{tabular}{@{}c@{}}$50$:$50$ beam splitter \\ \& photon counting\end{tabular} & $\pi/8$ quadratures \\
\hline
$1.44\cdot 10^{-1}$ & $1.49\cdot 10^{-1}$ & $1.47\cdot 10^{-1}$ & $9.94\cdot 10^{-2}$ & $1.57\cdot 10^{-1}$ & $1.27\cdot 10^{-1}$ \\
$1.11\cdot 10^{-1}$ & $1.15\cdot 10^{-1}$ & $1.13\cdot 10^{-1}$ & $6.48\cdot 10^{-2}$ & $1.23\cdot 10^{-1}$ & $8.37\cdot 10^{-2}$ \\
$8.94\cdot 10^{-2}$ & $9.25\cdot 10^{-2}$ & $9.07\cdot 10^{-2}$ & $4.49\cdot 10^{-2}$ & $9.71\cdot 10^{-2}$ & $5.83\cdot 10^{-2}$ \\
$7.47\cdot 10^{-2}$ & $7.70\cdot 10^{-2}$ & $7.56\cdot 10^{-2}$ & $3.36\cdot 10^{-2}$ & $7.85\cdot 10^{-2}$ & $4.31\cdot 10^{-2}$ \\
$6.40\cdot 10^{-2}$ & $6.59\cdot 10^{-2}$ & $6.47\cdot 10^{-2}$ & $2.64\cdot 10^{-2}$ & $6.44\cdot 10^{-2}$ & $3.32\cdot 10^{-2}$ \\
$5.60\cdot 10^{-2}$ & $5.74\cdot 10^{-2}$ & $5.66\cdot 10^{-2}$ & $2.17\cdot 10^{-2}$ & $5.38\cdot 10^{-2}$ & $2.67\cdot 10^{-2}$ \\
$4.98\cdot 10^{-2}$ & $5.10\cdot 10^{-2}$ & $5.02\cdot 10^{-2}$ & $1.83\cdot 10^{-2}$ & $4.56\cdot 10^{-2}$ & $2.22\cdot 10^{-2}$ \\
$4.48\cdot 10^{-2}$ & $4.58\cdot 10^{-2}$ & $4.51\cdot 10^{-2}$ & $1.58\cdot 10^{-2}$ & $3.91\cdot 10^{-2}$ & $1.89\cdot 10^{-2}$ \\
$4.07\cdot 10^{-2}$ & $4.15\cdot 10^{-2}$ & $4.10\cdot 10^{-2}$ & $1.40\cdot 10^{-2}$ & $3.39\cdot 10^{-2}$ & $1.64\cdot 10^{-2}$ \\
$3.74\cdot 10^{-2}$ & $3.80\cdot 10^{-2}$ & $3.76\cdot 10^{-2}$ & $1.25\cdot 10^{-2}$ & $2.98\cdot 10^{-2}$ & $1.45\cdot 10^{-2}$ \\
\hline
\hline
\multicolumn{3}{|c|}{Squeezed entangled state} & \multicolumn{3}{c|}{Twin squeezed cat state} \\
\hline
\begin{tabular}{@{}c@{}}Single-shot \\ measurement\end{tabular} & \begin{tabular}{@{}c@{}}$50$:$50$ beam splitter \\ \& photon counting\end{tabular} & $\pi/8$ quadratures & \begin{tabular}{@{}c@{}}Single-shot \\ measurement\end{tabular} & \begin{tabular}{@{}c@{}}$50$:$50$ beam splitter \\ \& photon counting\end{tabular} & $\pi/8$ quadratures \\
\hline 
$1.12\cdot 10^{-1}$ & $1.93\cdot 10^{-1}$ & $1.54\cdot 10^{-1}$ & $1.43\cdot 10^{-1}$ & $1.76\cdot 10^{-1}$ & $1.62\cdot 10^{-1}$ \\
$7.38\cdot 10^{-2}$ & $1.80\cdot 10^{-1}$ & $1.18\cdot 10^{-1}$ & $1.08\cdot 10^{-1}$ & $1.52\cdot 10^{-1}$ & $1.30\cdot 10^{-1}$ \\
$5.08\cdot 10^{-2}$ & $1.68\cdot 10^{-1}$ & $9.23\cdot 10^{-2}$ & $8.46\cdot 10^{-2}$ & $1.32\cdot 10^{-1}$ & $1.06\cdot 10^{-1}$ \\
$3.60\cdot 10^{-2}$ & $1.56\cdot 10^{-1}$ & $7.34\cdot 10^{-2}$ & $6.77\cdot 10^{-2}$ & $1.16\cdot 10^{-1}$ & $8.75\cdot 10^{-2}$ \\
$2.62\cdot 10^{-2}$ & $1.45\cdot 10^{-1}$ & $5.95\cdot 10^{-2}$ & $5.53\cdot 10^{-2}$ & $1.02\cdot 10^{-1}$ & $7.33\cdot 10^{-2}$ \\
$1.96\cdot 10^{-2}$ & $1.34\cdot 10^{-1}$ & $4.87\cdot 10^{-2}$ & $4.56\cdot 10^{-2}$ & $9.08\cdot 10^{-2}$ & $6.22\cdot 10^{-2}$ \\
$1.51\cdot 10^{-2}$ & $1.24\cdot 10^{-1}$ & $4.06\cdot 10^{-2}$ & $3.81\cdot 10^{-2}$ & $8.13\cdot 10^{-2}$ & $5.33\cdot 10^{-2}$ \\
$1.19\cdot 10^{-2}$ & $1.15\cdot 10^{-1}$ & $3.43\cdot 10^{-2}$ & $3.19\cdot 10^{-2}$ & $7.33\cdot 10^{-2}$ & $4.60\cdot 10^{-2}$ \\
$9.65\cdot 10^{-3}$ & $1.06\cdot 10^{-1}$ & $2.94\cdot 10^{-2}$ & $2.70\cdot 10^{-2}$ & $6.65\cdot 10^{-2}$ & $4.01\cdot 10^{-2}$ \\
$8.04\cdot 10^{-3}$ & $9.77\cdot 10^{-2}$ & $2.54\cdot 10^{-2}$ & $2.30\cdot 10^{-2}$ & $6.07\cdot 10^{-2}$ & $3.52\cdot 10^{-2}$ \\
\hline
\end{tabular}}
\caption{Mean square error for the indefinite photon number states using the optimal single-shot POVM and the physical measurement schemes described in the main text, with $1\leqslant\mu\leqslant 10$, $\bar{n}=2$, $\bar{\theta}=0$ and $W_0=\pi/2$.}
\label{practical_povm_summary}
\end{table*}

This calculation is shown in figure \ref{sinapprox_plot}, where the graph in the middle of the shaded areas is $\bar{\epsilon}_{\mathrm{mse}}$ for $1\leqslant\mu \leqslant 10$ and $W_0 = \pi/2$ and the boundaries are given by $\pm \Delta \bar{\epsilon}$. We can see that the Taylor error bounds for the twin squeezed cat state, the squeezed entangled state and the twin squeezed cat state, which constitute the basis of our main results, do not overlap for any value of $\mu$. Therefore, all the comparisons made between these probes are valid. That the twin squeezed cat state and the coherent state overlap for $\mu = 1, 2, 3$ is not surprising, since their respective mean square errors also do (see figure \ref{bounds_results}.i), and the same observation hold for the NOON state and the squeezed entangled state when $\mu = 2$. On the other hand, the shaded area of the NOON state overlaps slightly with the top shaded area of the twin squeezed vacuum state when $\mu=1$. It is important to appreciate that the shaded areas are bounds for the Taylor error, and it is not guaranteed that the uncertainty for this two states actually coincides. However, even if they did, it would simply constitute another instance where the role of inter-mode and intra-mode correlations is altered in the regime of limited data, since a state with path entanglement that is beaten by a state with a large amount of intra-mode correlations in the asymptotic regime would reach the same uncertainty than the latter for a single shot. 

Finally, we also notice that the approximation will become even better as $W_0$ decreases, which is the case for the other prior widths that we have explored. Hence, we can conclude that the results that arise from the use of the mean square error as an approximation for the mean sine error in the regime of moderate prior knowledge are valid.

\section{Calculation scheme for the optimal single-shot strategy and its bound}\label{numcal}

\begin{table*} [t]
{\renewcommand{\arraystretch}{1.2} 
\begin{tabular}{|c|c|c|c|c|}
\hline 
\multicolumn{5}{|c|}{$\bar{\epsilon}_{\mathrm{mse}}\left(\mu = 1\right)$, $\dots$, $\bar{\epsilon}_{\mathrm{mse}}\left(\mu = 10 \right)$} \\
\hline
\hline
\multicolumn{5}{|c|}{NOON state} \\
\hline
\begin{tabular}{@{}c@{}}Single-shot \\ measurement\end{tabular} & \begin{tabular}{@{}c@{}}$50$:$50$ beam splitter \\ \& photon counting\end{tabular} & $\pi/8$ quadratures  & Parity measurements & Collective measurements\\
\hline 
$1.04\cdot 10^{-1}$ & $1.04\cdot 10^{-1}$ & $1.04\cdot 10^{-1}$ & $1.04\cdot 10^{-1}$ & $1.04\cdot 10^{-1}$ \\
$7.06\cdot 10^{-2}$ & $7.06\cdot 10^{-2}$ & $7.06\cdot 10^{-2}$ & $7.06\cdot 10^{-2}$ & $7.02\cdot 10^{-2}$ \\
$5.36\cdot 10^{-2}$ & $5.36\cdot 10^{-2}$ & $5.36\cdot 10^{-2}$ & $5.35\cdot 10^{-2}$ & $5.31\cdot 10^{-2}$ \\
$4.33\cdot 10^{-2}$ & $4.33\cdot 10^{-2}$ & $4.33\cdot 10^{-2}$ & $4.32\cdot 10^{-2}$ & $4.28\cdot 10^{-2}$ \\
$3.63\cdot 10^{-2}$ & $3.63\cdot 10^{-2}$ & $3.63\cdot 10^{-2}$ & $3.63\cdot 10^{-2}$ & $3.59\cdot 10^{-2}$ \\
$3.14\cdot 10^{-2}$ & $3.13\cdot 10^{-2}$ & $3.13\cdot 10^{-2}$ & $3.13\cdot 10^{-2}$ & $3.09\cdot 10^{-2}$ \\
$2.76\cdot 10^{-2}$ & $2.76\cdot 10^{-2}$ & $2.76\cdot 10^{-2}$ & $2.76\cdot 10^{-2}$ & $2.72\cdot 10^{-2}$ \\
$2.46\cdot 10^{-2}$ & $2.46\cdot 10^{-2}$ & $2.46\cdot 10^{-2}$ & $2.46\cdot 10^{-2}$ & $2.43\cdot 10^{-2}$ \\
$2.23\cdot 10^{-2}$ & $2.23\cdot 10^{-2}$ & $2.23\cdot 10^{-2}$ & $2.23\cdot 10^{-2}$ & $2.20\cdot 10^{-2}$ \\
$2.03\cdot 10^{-2}$ & $2.03\cdot 10^{-2}$ & $2.03\cdot 10^{-2}$ & $2.03\cdot 10^{-2}$ & $2.00\cdot 10^{-2}$ \\ 
\hline
\end{tabular}}
\caption{Mean square error for the NOON state using the optimal single-shot POVM, the physical measurement schemes described in the main text and collective measurements, with $1\leqslant\mu\leqslant 10$, $\bar{n}=2$, $\bar{\theta}=0$ and $W_0=\pi/2$. We note that the calculation for collective measurements has been performed with a different numerical algorithm.}
\label{noon_povm_summary}
\end{table*}

In this appendix we present the calculation scheme that has been used to obtain the eigenvalues and eigenvectors of the optimal quantum estimator $S$ that generate the results of the main text. 

We recall that the Hermitian operator $S$ satisfies the equation $S \rho + \rho S = 2 \bar{\rho}$. Hence, the first step is to find $\rho$ and $\bar{\rho}$. By expanding the transformed state $\ket{\psi(\theta)}$ in the number basis as $\ket{\psi(\theta)}= \sum_{nm}\mathrm{e}^{-i (n-m)\theta/2}c_{nm}\ket{nm}$, where $c_{nm}$ are the components of the initial probe state, we can construct the density matrix
\begin{equation}
\rho(\theta) = \sum_{nmlk} \mathrm{e}^{-i (n-m)\theta/2} \mathrm{e}^{i (k-l)\theta/2} c_{nm}c^{*}_{kl}\ketbra{nm}{kl},
\end{equation}
with $c_{nm}c^{*}_{kl} = (\rho_0)_{nmkl}$. Then, given the flat prior in equation (\ref{prior_probability}) we have that
\begin{equation}
\rho = \int d\theta p(\theta) \rho(\theta)= \sum_{nmkl} K_{nmkl} c_{nm}c^{*}_{kl}\ketbra{nm}{kl}
\label{zeroth_qmoment_code}
\end{equation}
and
\begin{equation}
\bar{\rho} = \int d\theta p(\theta)\rho(\theta) \theta = \sum_{nmkl} L_{nmkl} c_{nm}c^{*}_{kl}\ketbra{nm}{kl},
\label{first_qmoment_code}
\end{equation}
where
\begin{eqnarray}
K_{nmkl} = \frac{1}{W_0}\int_{\bar{\theta}-W_0/2}^{\bar{\theta}+W_0/2}d\theta \mathrm{e}^{-ix_{nmkl}\theta/2},
\label{k_semianalytical_int}
\end{eqnarray}
\begin{eqnarray}
L_{nmlk} &=& \frac{1}{W_0}\int_{\bar{\theta}-W_0/2}^{\bar{\theta}+W_0/2} d\theta \theta \mathrm{e}^{-ix_{nmkl}\theta/2}
\end{eqnarray}
and $x_{nmkl}=n-m+l-k$. These integrals can be computed directly, finding that
\begin{eqnarray}
K_{nmkl} = \frac{4}{W_0}\frac{A_{nmkl} B_{nmkl}}{x_{nmkl}}
\label{k_semianalytical_sol}
\end{eqnarray}
and
\begin{eqnarray}
L_{nmkl} &=& \frac{2 A_{nmkl}}{x_{nmkl}} \left(\frac{2 B_{nmkl} D_{nmkl}}{W_0} + iC_{nmkl}\right),
\end{eqnarray}
where we have defined
\begin{eqnarray}
A_{nmkl}&=&\mathrm{exp}\left(-i x_{nmkl}\bar{\theta}/2\right),
\nonumber \\
B_{nmkl}&=&\mathrm{sin}\left(x_{nmkl}W_0/4\right),
\nonumber \\
C_{nmkl}&=&\mathrm{cos}\left(x_{nmkl}W_0/4\right)~\mathrm{and}~
\nonumber \\
D_{nmkl}&=&\bar{\theta}-2 i/x_{nmkl}.
\label{d_semianalytical_auxiliar}
\end{eqnarray}
Note that all the elements $K_{nmkl}$ and $L_{nmkl}$ are well defined except when $x_{nmkl}$ vanishes, in which case we have an indetermination. In those cases we need to take the limits
\begin{equation}
\underset{x_{nmlk}\rightarrow 0}{\mathrm{lim}} K_{nmkl}=1 ,~~\underset{x_{nmkl}\rightarrow 0}{\mathrm{lim}} L_{nmkl}=\bar{\theta}.
\end{equation}

Since $K_{nmkl}$, $L_{nmkl}$ and $c_{nm}c^{*}_{kl}$ can be seen as $(nm \times kl)$ matrices, we can rewrite equations (\ref{zeroth_qmoment_code}) and (\ref{first_qmoment_code}) as $\rho = \rho_0 \circ K$ and $\rho = \rho_0 \circ L$, respectively, where we are using the entrywise product of matrices defined as $X \circ Y = \sum_{ij} X_{ij} Y_{ij}\ketbra{i}{j}$ \cite{horn1985}. In other words, now we have two expressions where the integration has been performed analytically.

On the other hand, if we expand $\rho$ in the basis of its eigenvectors, that is, $\rho = \sum_i p_i \ketbra{\phi_i}$, and we insert it into $S \rho + \rho S = 2 \bar{\rho}$, we arrive to 
\begin{equation}
S = 2\sum_{ij} \frac{\bra{\phi_i} \bar{\rho} \ket{\phi_j}}{p_i+p_j}\ketbra{\phi_i}{\phi_j}.
\label{quantumestimator_rho}
\end{equation}
Note that (\ref{quantumestimator_rho}) is only defined on the support of $\rho$, since the Sylvester equation $S \rho + \rho S = 2 \bar{\rho}$ only has a unique solution in the subspace where the spectra of $\rho$ and $-\rho$ are disjoint \cite{bhatia1997}. Interestingly, this solution is formally analogous to the expression to calculate the symmetric logarithmic derivative in the asymptotic theory \cite{paris2009, rafal2015}.

Unfortunately, completing this calculation analytically for the indefinite photon number states is challenging because they belong to a Hilbert space whose dimension is infinite. However, we can take advantage of the analytical expressions $\rho = \rho_0 \circ K$, $\rho = \rho_0 \circ L$ and those in equations (\ref{k_semianalytical_sol}) - (\ref{quantumestimator_rho}) in order to simplify the numerical scheme. In particular, we have implemented the following method:
\begin{enumerate}
\item The components $c_{nm}$ of the initial state $\ket{\psi_0}$ are numerically approximated employing a finite Hilbert space of dimension $d_c$ per mode. For the coherent state this dimension is $d_c=21$, and the number probability for this cut-off is $p_c \sim 10^{-19}$; for the twin squeezed vacuum state we have that $d_c = 51$ and $p_c \sim 10^{-17}$; $d_c = 61$ and $p_c \sim 10^{-5}$ for the squeezed entangled state; and $d_c = 51$ and $p_c \sim 10^{-10}$ for the twin squeezed cat state.
\item The matrices $K$ and $L$ are numerically generated using the formulas in equations (\ref{k_semianalytical_sol}) - (\ref{d_semianalytical_auxiliar}). This allows us to calculate $\rho = \rho_0 \circ K$ and $\rho = \rho_0 \circ L$ in the number basis.
\item The basis of $\rho$ and $\bar{\rho}$ is changed as $\rho_D = V^\dagger \rho V$ and $\bar{\rho}_D = V^\dagger \bar{\rho} V$, where the columns of $V$ are given by the eigenvectors $\ket{\phi_i}$ of $\rho$, $(\rho_D)_{ij} = p_i\delta_{ij}$ and $(\bar{\rho}_D)_{ij} = \bra{\phi_i} \bar{\rho} \ket{\phi_j}$.
We note that only the eigenvectors $\ket{\phi_i}$ whose eigenvalues $p_i$ satisfy that $p_i \gtrsim 10^{-12}$ are employed. 
\item Now we can calculate the elements $(S_D)_{ij} = \bra{\phi_i} S \ket{\phi_j}=2(\bar{\rho}_D)_{ij}/(p_i + p_j)$ directly.
\item We return to the original basis using $S=V S_D V^\dagger$.
\item Finally, we calculate the spectral decomposition of $S$ as indicated in equation (\ref{singleshot_strategy}), which gives us the estimates $\lbrace s\rbrace$ and the projectors $\lbrace \ket{s}\rbrace$.
\end{enumerate}

Once the optimal single-shot POVM has been found, we can proceed with the calculation of the mean square error as a function of $\mu$ in equation (\ref{mse_mu_repetitions}) using the three-step method in \cite{jesus2017}. Tables \ref{practical_povm_summary} and \ref{noon_povm_summary} provide the numerical values of our schemes for $1 \leqslant \mu \leqslant 10$, while the complete results for $1 \leqslant \mu \leqslant 10^3$ have been presented as graphs in the main text. The numerical precision of these values can be estimated using the identity
\begin{equation}
\int d\theta p(\theta) \theta^2 = \int d\theta' p(\theta') \int d\boldsymbol{n}~ p(\boldsymbol{n}|\theta') \int d\theta p(\theta|\boldsymbol{n})\theta^2,
\end{equation}
where the right hand side is calculated numerically and it is compared to the analytical solution for the left hand side. In particular, we have found that our results are valid up to the third significant figure.  

\section{Analytical results for the NOON state}\label{noon_analytical}

Given a NOON state with $\bar{n}=2$ photons, the transformed probe after encoding the unknown parameter is
\begin{eqnarray}
\ket{\psi(\theta)} &=& \frac{1}{\sqrt{2}}\left(\mathrm{e}^{-i\theta}\ket{2,0} + \mathrm{e}^{i\theta}\ket{0,2} \right).
\end{eqnarray}
In that case, we have that
\begin{equation}
\rho = \frac{2}{\pi}\int_{-\pi/4}^{\pi/4} d\theta \ketbra{\psi(\theta)} = \frac{\mathbb{I}}{2} + \frac{\sigma_x}{\pi}
\label{rhonoon}
\end{equation}
and
\begin{equation}
\bar{\rho} = \frac{2}{\pi}\int_{-\pi/4}^{\pi/4} d\theta \ketbra{\psi(\theta)} \theta = \frac{\sigma_y}{2\pi},
\label{rhobarnoon}
\end{equation}
where $\sigma_x$ and $\sigma_y$ are Pauli matrices. Inserting equations (\ref{rhonoon}) and (\ref{rhobarnoon}) into $S \rho + \rho S = 2 \bar{\rho}$ we find that the quantum estimator is $S = \sigma_y/\pi$. Hence, the optimal single-shot POVM that we discussed in section (\ref{main_results}) is given by the eigenvectors
\begin{equation}
\ket{s_1} = \frac{1}{\sqrt{2}} 
\begin{pmatrix} 
i  \\
1  
\end{pmatrix}, ~~
\ket{s_2} = \frac{1}{\sqrt{2}}
\begin{pmatrix} 
 1 \\
i 
\end{pmatrix},
\end{equation}
which are associated with the Bayesian estimates $-1/\pi$ and $1/\pi$, respectively. These eigenvalues were represented in figure \ref{bayes_spectra}.

Next we calculate the optimal mean square error for a single shot using the bound in equation (\ref{singleshot_bound}), finding that
\begin{equation}
\bar{\epsilon}_{\mathrm{mse}}(\mu=1) \geqslant \frac{\pi^2}{48} - \frac{1}{\pi^2}  \approx 0.104,
\label{mse_analytical_noon}
\end{equation}
which is in perfect agreement with the numerical results showed in table \ref{noon_povm_summary}.

In section \ref{measurements_section} we demonstrated that several physical measurements were able to saturate equation (\ref{mse_analytical_noon}) for the NOON state using numerical techniques. Here we will recover the same result analytically for one of them. In particular, let us consider the POVM based on counting photons as indicated by Table \ref{povm_summary}. First we construct the likelihood function 
\begin{equation}
p(n,m|\theta) = ||\bra{n,m} \mathrm{e}^{-i\frac{\pi}{2}J_x} \mathrm{e}^{-i\frac{\pi}{4}N_2} \ket{\psi(\theta)}||^2,
\end{equation}
where $N_2 = a^\dagger_2 a_2$, and after manipulating and simplifying this expression, we arrive to
\begin{eqnarray}
p(2,0|\theta) &=& p(0,2|\theta) = \frac{1}{2}~\mathrm{sin}^2\left(\theta - \frac{\pi}{4}\right),
\nonumber \\
p(1,1|\theta) &=& \mathrm{cos}^2\left(\theta - \frac{\pi}{4}\right).
\label{likelihood_noon}
\end{eqnarray}
It is interesting to observe that the width of the ranges where $p(n,m|\theta)$ is monotonic is $\pi/2$, and according to \cite{kolodynski2014}, this is precisely the maximum value that we can assign to $W_0$ while still being able to infer the unknown parameter unambiguously, an idea that was captured by the concept of intrinsic width in \cite{jesus2017}. This is why the NOON state can also be useful in the regime of moderate prior knowledge.

The next quantity that we need to find is the normalisation term of Bayes' theorem, that is,
\begin{equation}
p(n,m) = \frac{2}{\pi} \int_{-\pi/4}^{\pi/4} d\theta p(n,m|\theta).
\end{equation}
Introducing equation (\ref{likelihood_noon}) into the formula for $p(n,m)$, we find that $p(2,0) = p(0,2) = 1/4$ and $p(1,1)=1/2$. At the same time, this gives us the last piece that we need to apply Bayes' theorem and find the posterior probability $p(\theta|n,m) = p(\theta)p(n,m|\theta)/p(n,m)$, which in our case is
\begin{eqnarray}
p(\theta|2,0) &=& p(\theta|0,2) = \frac{4}{\pi}~\mathrm{sin}^2\left(\theta - \frac{\pi}{4}\right),
\nonumber \\
p(\theta|1,1) &=& \frac{4}{\pi}~\mathrm{cos}^2\left(\theta - \frac{\pi}{4}\right).
\label{posterior_noon}
\end{eqnarray}

On the other hand, it is possible to rewrite the single-shot mean square error in equation (\ref{singleshotmse}) as
\begin{equation}
\bar{\epsilon}_{\mathrm{mse}}(\mu=1) = \int d\theta p(\theta) \theta^2 - \int dn~ p(n,m) g_{\mathrm{opt}}(n,m)^2,
\end{equation}
where
\begin{eqnarray}
g_{\mathrm{opt}}(n,m) &=& \int_{-\pi/4}^{\pi/4} d\theta p(\theta|n,m)\theta
\end{eqnarray}
is the optimal classical estimator. Taking into account that $g_{\mathrm{opt}}(2,0) = g_{\mathrm{opt}}(0,2) = -1/\pi$ and $g_{\mathrm{opt}}(1,1) = 1/\pi$, the error associated to this POVM is
\begin{eqnarray}
\bar{\epsilon}_{\mathrm{mse}}(\mu=1) &=& \frac{2}{\pi}\int_{-\pi/4}^{\pi/4} d\theta \theta^2 - \frac{1}{\pi^2} 
\nonumber \\
&=& \frac{\pi^2}{48} - \frac{1}{\pi^2},
\end{eqnarray}
which, as expected, saturates the bound in equation (\ref{mse_analytical_noon}).

\section{Optimal single-shot mean square error in the limit of a narrow flat prior}\label{prior_appendix}

Given the uniform prior probability introduced in equation (\ref{prior_probability}), and assuming that $W_0\ll 1$, the Taylor expansion around $\bar{\theta}$ for the transformed state $\rho(\theta)$ is
\begin{equation} 
\rho(\theta)  \approx \rho(\bar{\theta}) + \frac{\partial \rho(\bar{\theta})}{\partial \theta} \left(\theta - \bar{\theta} \right).
\end{equation}
Furthermore, recalling that the symmetric logarithmic derivative is defined as \cite{helstrom1976,paris2009,rafal2015}
\begin{equation}
\frac{\partial \rho(\bar{\theta})}{\partial \theta} = \frac{1}{2}\left[ L(\bar{\theta})\rho(\bar{\theta}) + \rho(\bar{\theta}) L(\bar{\theta})\right],
\end{equation}
we have that 
\begin{equation}
\rho(\theta)  \approx \rho(\bar{\theta}) + \frac{1}{2}\left[ L(\bar{\theta})\rho(\bar{\theta}) + \rho(\bar{\theta}) L(\bar{\theta})\right] \left(\theta - \bar{\theta} \right).
\label{density_matrix_approx}
\end{equation}

The next step is to introduce equation (\ref{density_matrix_approx}) into $\rho = \int d\theta p(\theta) \rho(\theta)$ and $\bar{\rho} = \int d\theta p(\theta) \rho(\theta) \theta$, finding that $\rho \approx \rho(\bar{\theta})$ and
\begin{equation}
\bar{\rho} \approx \bar{\theta}\rho(\bar{\theta}) +  \frac{\Delta \theta^2_p}{2}\left[ L(\bar{\theta})\rho(\bar{\theta}) + \rho(\theta) L(\bar{\theta})\right].
\end{equation}
Thus, using equation $S\rho + \rho S = 2\bar{\rho}$ we can see that the quantum estimator takes the form $S \approx \bar{\theta} ~\mathbb{I} + \Delta \theta^2_p ~ L(\bar{\theta})$, and this implies that $\mathrm{Tr}\left(\bar{\rho}S\right) \approx \bar{\theta}^2 + \Delta \theta^4_p~F_q(\bar{\theta})$, where $F_q(\bar{\theta}) = \mathrm{Tr}[\rho(\bar{\theta})L(\bar{\theta})^2]$ is the quantum Fisher information and we have used the fact that $\mathrm{Tr}\left[\rho(\bar{\theta})L(\bar{\theta})\right] = 0$.

Finally, we arrive to
\begin{equation}
\bar{\epsilon}_{\mathrm{mse}} \gtrsim \Delta \theta^2_p \left[1-\Delta \theta^2_p F_q(\bar{\theta}) \right]
\label{narrowprior_approx}
\end{equation}
after introducing the approximated expression for $\mathrm{Tr}\left(\bar{\rho}S\right)$ into equation (\ref{singleshot_bound}), which is the result involved in the discussion of Section \ref{prior_section} and that was available in the literature for a Gaussian prior \cite{macieszczak2014bayesian, jarzyna2015true, jarzyna2016thesis}.

\section{Calculation scheme for the optimal mean square error with two identical probes}\label{numcal_noon}

The calculation of equation (\ref{singleshot_collective}) for collective measurements is completely analogous to the scheme developed in appendix \ref{numcal} for $\mu = 1$. For instance, when the number of copies is $\mu = 2$, equations (\ref{zerothmoment_collective}) and (\ref{firstmoment_collective}) can be expressed as
\begin{eqnarray}
\rho = \sum_{\substack{nmlk \\ abcd}} L_{n m k l a b c d}  c_{nm}c^{*}_{kl}c_{ab}c^{*}_{cd} \ketbra{nmab}{klcd}
\end{eqnarray}
and
\begin{eqnarray}
\bar{\rho}  = \sum_{\substack{nmlk \\ abcd}} K_{n m k l a b c d}  c_{nm}c^{*}_{kl}c_{ab}c^{*}_{cd} \ketbra{nmab}{klcd},
\end{eqnarray}
where $L_{n m k l a b c d}$ and $K_{n m k l a b c d}$ are defined in the same way as in equations (\ref{k_semianalytical_int}) - (\ref{d_semianalytical_auxiliar}), but using 
\begin{equation}
y_{n m k l a b c d} = n-m+l-k+a-b+d-c
\end{equation}
instead of $x_{n m k l}$. The same principle can be trivially generalised to larger numbers of copies. Nevertheless, it becomes numerically challenging to create objects with such a large amount of indices, which is why we have only considered $1\leqslant \mu \leqslant 10$ for the results based on collective measurements. The last column of table \ref{noon_povm_summary} and the graph of figure \ref{noonpovm}.iv have been generated using this method.

\section{The effect of photon losses}
\label{loss}

 While a comprehensive analysis of the combined effect of having some form of noise and a limited amount of data is left for future research, in this appendix we present an initial test of the type of behaviour that we could expect from the application of our method to noisy scenarios. 

Following \cite{dorner2009}, suppose we consider another Mach-Zehnder interferometer with initial state $\ket{\psi_0} = \sum_{k=0}^2 c_k \ket{k,2-k}$ and where the unknown phase shift $\phi$ is now encoded in the first arm with the unitary transformation $\mathrm{exp}(-i N_1 \phi)$, where $N_i=a_1^\dagger a_1$. In addition, the photon losses in such arm are modelled using a fictitious beam splitter with transmissivity $\eta$. In that case, the transformed state is \cite{dorner2009}
\begin{equation}
\rho(\phi) = \mathrm{e}^{-i N_1 \phi}\left(\sum_{l=0}^2 K_{l,a_1}\ketbra{\psi_0}K_{l,a_1}^\dagger\right) \mathrm{e}^{i N_1 \phi},
\label{lossy_state}
\end{equation}
where $K_{k,a_1}=(1-\eta)^{l/2}\eta^{N_1/2}a_1^l/\sqrt{l!}$ are Krauss operators.

We need to find the state $\ket{\psi_0}$ that is optimal for a given amount of loss. Since for this initial test we are interested in analysing the specific proposal in \cite{dorner2009} and this work is based on the Fisher information, we will simply select the initial probe that has the largest $F_q$, and we will follow the methodology in the main text to find the Bayesian bound based on repeating the optimal single-shot strategy of this state. However, a potentially better result could be found by optimising the single-shot bound instead. We leave this possibility for future work.

To represent a realistic amount of loss we can choose $\eta = 9/10$, and the components of the state with the largest $F_q$ for this value are $c_0 = 3/\sqrt{10}$, $c_1 = 0$ and $c_2 = \sqrt{10/19}$. Hence, equation (\ref{lossy_state}) becomes
\begin{equation}
\rho(\phi) = \frac{1}{190} 
\begin{pmatrix} 
1 & 0 & 0 & 0 \\
0 & 90 & 0 & 27 \sqrt{10}~\mathrm{e}^{i 2 \phi} \\
0 & 0 & 18 & 0 \\
0 & 27 \sqrt{10}~\mathrm{e}^{-i 2 \phi} & 0 & 81 
\end{pmatrix},
\end{equation}
where the columns are labelled as $\ket{0,0}$, $\ket{0,2}$, $\ket{1,0}$ and $\ket{2,0}$, respectively.

\begin{figure}[t]
\centering
\includegraphics[trim={0.2cm 0.1cm 1.3cm 0.5cm},clip,width=9cm]{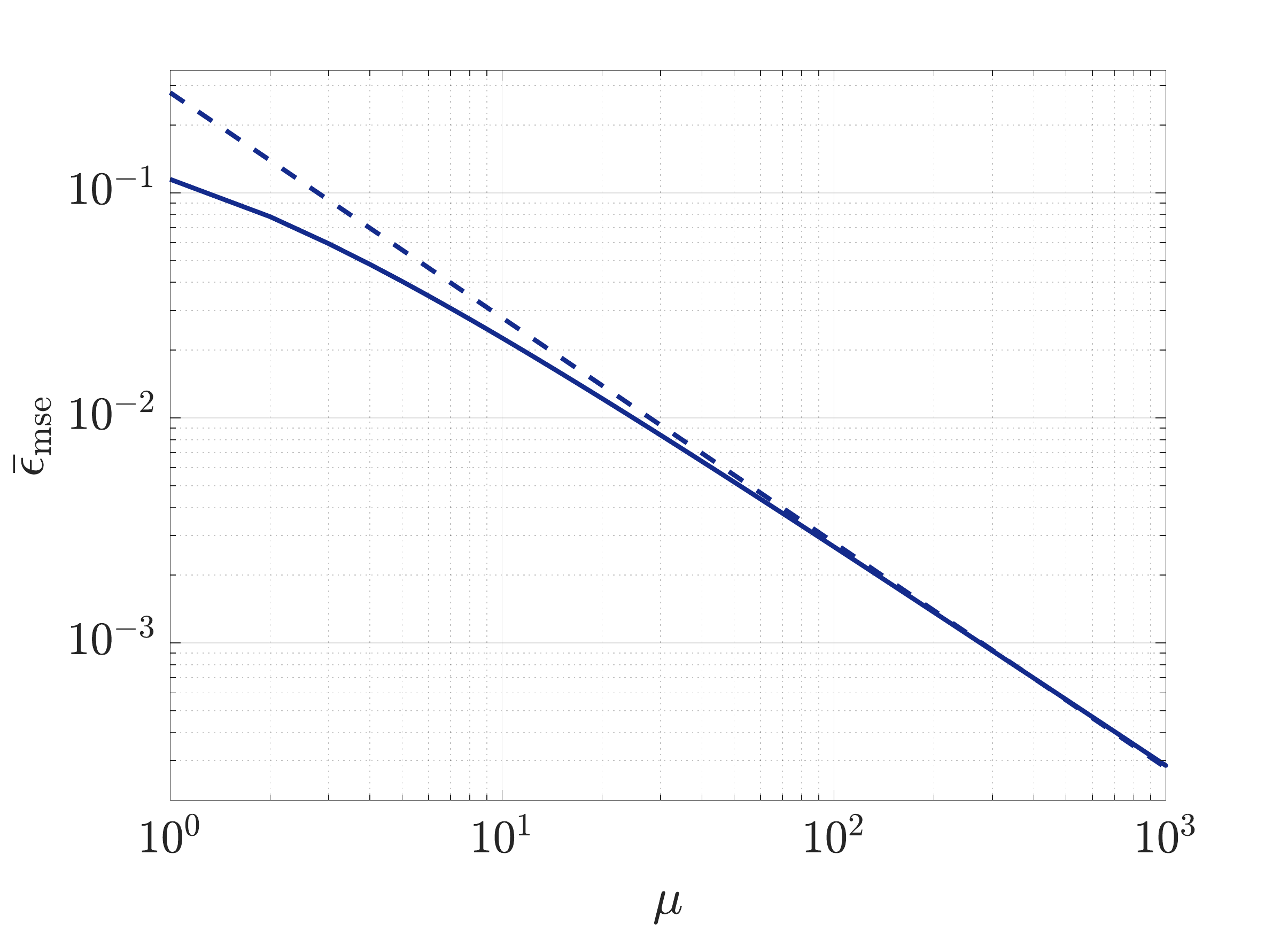}
	\caption{Mean square error based on the optimal single-shot strategy (solid line) and quantum Cram\'{e}r-Rao bound (dashed line) for a two-photon state whose Fisher information is optimal (see \cite{dorner2009}) that is fed to a Mach-Zehnder interferometer with photon losses in its first arm, with $\eta=0.9$, $\bar{\phi} = \pi/4$ and $W_0 = \pi/2$.}
\label{lossy_plot}
\end{figure}

The next step is to calculate the optimal single-shot strategy. Assuming that the prior $p(\phi)$ is a flat density of width $W_0 = \pi/2$ and centred around $\bar{\phi} = \pi/4$, we can calculate $\rho = \int d\phi p(\phi)\rho(\phi)$, $\bar{\rho} = \int d\phi p(\phi)\rho(\phi)\phi$ and insert the results into $S\rho + \rho S = 2\bar{\rho}$ to find the optimal quantum estimator
\begin{equation}
S=\frac{1}{76\pi}
\begin{pmatrix} 
19\pi^2 & 0 & 0 & 0 \\
0 & 19\pi^2 & 0 & -24 \sqrt{10} \\
0 & 0 & 19\pi^2 & 0 \\
0 & -24 \sqrt{10} & 0 & 19\pi^2
\end{pmatrix}
\end{equation}
whose eigenspaces allow us to construct the projective measurement $\ket{s_1}=(-\ket{0,2}+\ket{2,0})/\sqrt{2}$, $\ket{s_2}=(\ket{0,2}+\ket{2,0})/\sqrt{2}$, $\ket{s_3}=\ket{1,0}$ and $\ket{s_4}=\ket{0,0}$. Note that in this case the optimal single-shot POVM is not unique due to the degeneracy of one of the eigenvalues of $S$.

Finally, we calculate the mean square error in equation (\ref{mse_mu_repetitions}) using this optimal single-shot measurement. The result has been represented in figure \ref{lossy_plot}, which also includes the quantum Cram\'{e}r-Rao bound that can be obtained using the expression for the Fisher information provided in \cite{dorner2009}. As we can see, the Bayesian error approaches the asymptotic result also in this case, and while a perfect convergence cannot be observed within the number of trials that we are considering because the mean square error crosses the bound when $\mu \approx 4 \cdot 10^2$, we have verified that after $\mu = 10^3$ repetitions the relative error defined in equation (\ref{threshold}) is just $\varepsilon = 0.02$. Therefore, we can conclude that, according to our methodology, a reasonably amount of photon losses does not seem to alter substantially the behaviour that we have found in the main text using ideal schemes. Nevertheless, a deeper investigation including other sources of noise, other probe states and realistic measurements is required in order to construct a more complete picture of the effect that noise has in those systems that operate in the regime of limited data.

\end{document}